\documentclass[proc]{edpsmath}
\pdfoutput=0
\usepackage[english]{babel}

\usepackage{graphicx}

\usepackage{url}
\usepackage{tikz}

\usepackage{amsfonts}
\usepackage{mathrsfs}
\usepackage{amssymb}

\usepackage{mathtools}

\usepackage{algorithm2e}

\usepackage{ifthen}

\usepackage{moreverb}

\usepackage{ stmaryrd }

\def\Nr{{\rm N_r}}

\def\Ntheta{{\rm N_\theta}}

\newcommand{\gy}{\textsc{Gysela}\xspace}
\newcommand{\hel}{\textsc{Helios}\xspace}
\newcommand{\mpi}{\textsc{MPI}\xspace}
\newcommand{\omp}{\textsc{OpenMP}\xspace}

\definecolor{darkgreen}{rgb}{0.1, 0.45, 0.0}
\definecolor{lightgreen}{rgb}{0.5, 0.9, 0.2}

\newcommand{\rhsval}{$W_f$}

\ifthenelse{\boolean{true}}{
\newcommand{\todocomment}[1]   {\textbf{#1}}
\newcommand{\todo}[1]          {{\color{red}\textbf{TODO:} #1}}

}{
\newcommand{\todocomment}[1]   {}
\newcommand{\todo}[1]          {}

}

\usetikzlibrary{matrix, shapes, arrows, calc, positioning, decorations.pathmorphing}

\newcommand{\cell}[2] {
  \fill [color = gray!20 ] ( #1 : #2 ) node [scale=0.7] {{\color{lightgreen}$\blacksquare$}}
  arc ( #1 : #1+15 : #2 ) node [scale=0.7] {{\color{lightgreen}$\blacksquare$}}
  --  ( #1+15 : #2 -1 ) node [scale=0.7] {{\color{lightgreen}$\blacksquare$}}
  arc ( #1+15 : #1 : #2 -1 ) node [scale=0.7] {{\color{lightgreen}$\blacksquare$}}
  -- cycle ;
}

\newcommand{\cellclear}[2] {
  \def\myscale{1.1}
  \draw ( #1 : #2 ) node [scale=\myscale] {{\color{lightgreen}$\blacksquare$}}
  arc ( #1 : #1+15 : #2 ) node [scale=\myscale] {{\color{lightgreen}$\blacksquare$}}
  --  ( #1+15 : #2 -1 ) node [scale=\myscale] {{\color{lightgreen}$\blacksquare$}}
  arc ( #1+15 : #1 : #2 -1 ) node [scale=\myscale] {{\color{lightgreen}$\blacksquare$}} 
  -- cycle ;
}

\newcommand{\hcell}[4] {
  \def\deltawidth{ #3/4 }
  \def\quarter{ #1+\deltawidth }
  \def\midcell{ #1+2*\deltawidth }
  \def\thirdquarter{ #1+3*\deltawidth }

  \draw [ thick ] (#1,#2) -- ( #1+#3,#2) -- (#1+#3,#2+#4) -- (#1,#2+#4) -- cycle;
  \draw ( \quarter, #2 ) -- ( \quarter, #2+#4 );
  \draw ( \midcell, #2 ) -- ( \midcell, #2+#4 );
  \draw ( \thirdquarter, #2 ) -- ( \thirdquarter, #2+#4 );
}

\newcommand{\emptyhcell}[4] {
  \draw [ thick ] (#1,#2) -- ( #1+#3,#2) -- (#1+#3,#2+#4) -- (#1,#2+#4) -- cycle;
}

\newcommand{\vcell}[4] {
  \def\deltaheight{ #4/4 }
  \def\quarter{ #2+\deltaheight }
  \def\midcell{ #2+2*\deltaheight }
  \def\thirdquarter{ #2+3*\deltaheight }

  \draw [ thick ] (#1,#2) -- ( #1+#3,#2) -- (#1+#3,#2+#4) -- (#1,#2+#4) -- cycle;
  \draw ( #1, \quarter ) -- ( #1+#3, \quarter );
  \draw ( #1, \midcell ) -- ( #1+#3, \midcell );
  \draw ( #1, \thirdquarter ) -- ( #1+#3, \thirdquarter );
}

\newcommand{\filledhcell}[4] {
  \def\deltawidth{ #3/4 }
  \def\quarter{ #1+\deltawidth }
  \def\midcell{ #1+2*\deltawidth }
  \def\thirdquarter{ #1+3*\deltawidth }

  \fill [ thick, color=gray!60 ] (#1,#2) -- ( #1+#3,#2) -- (#1+#3,#2+#4) -- (#1,#2+#4) -- cycle;
}

\newcommand{\dothcell}[4] {
  \def\deltawidth{ #3/2 }
  \def\midwidth{ #1 + \deltawidth }
  \def\deltaheight{ #4/2 }
  \def\midheight{ #2 + \deltaheight }

  \draw [ thick ] (#1,#2) -- ( #1+#3,#2) -- (#1+#3,#2+#4) -- (#1,#2+#4) -- cycle;
  \node at ( \midwidth, \midheight ) { $\ldots$ } ;
}

\newcommand{\dotvcell}[4] {
  \def\deltawidth{ #3/2 }
  \def\midwidth{ #1 + \deltawidth }
  \def\deltaheight{ #4/2 }
  \def\midheight{ #2 + \deltaheight }

  \draw [ thick ] (#1,#2) -- ( #1+#3,#2) -- (#1+#3,#2+#4) -- (#1,#2+#4) -- cycle;
  \node at ( \midwidth, \midheight ) { $\vdots$ } ;
}

\newcommand{\alphahcell}[6] {
  \node [ thin, scale = 1, draw ] (coef) at ( #1, #2 ) 
  {$\alpha_{i,j}^0$ $\alpha_{i,j}^1$ $\alpha_{i,j}^2$ $\alpha_{i,j}^3$};
  \draw [ thick ] (coef.north west) -- (coef.north east);
  \draw [ thick ] (coef.south west) -- (coef.south east);
  \draw [ thick ] (coef.south west) -- (coef.north west);
  \draw [ thick ] (coef.south east) -- (coef.north east);
  \draw [ thin ] (coef.154) -- (coef.206);
  \draw [ thin ] (coef.90) -- (coef.270);
  \draw [ thin ] (coef.26) -- (coef.334);

  \draw [ thin ] ( #3, #4 ) -- (coef.south west);
  \draw [ thin ] ( #5, #6 ) -- (coef.south east);
}

\newcommand{\fvalhcell}[6] {
  \node [ thin, scale = 1, draw ] (coef) at ( #1, #2 ) 
  {$f_{i,j}^0$ $f_{i,j}^1$ $f_{i,j}^2$ $f_{i,j}^3$};
  \draw [ thick ] (coef.north west) -- (coef.north east);
  \draw [ thick ] (coef.south west) -- (coef.south east);
  \draw [ thick ] (coef.south west) -- (coef.north west);
  \draw [ thick ] (coef.south east) -- (coef.north east);
  \draw [ thin ] (coef.151) -- (coef.209);
  \draw [ thin ] (coef.90) -- (coef.270);
  \draw [ thin ] (coef.29) -- (coef.331);

  \draw [ thin ] ( #3, #4 ) -- (coef.south west);
  \draw [ thin ] ( #5, #6 ) -- (coef.south east);
}

\newcommand{\nextfvalhcell}[6] {
  \node [ thin, scale = 1, draw ] (coef) at ( #1, #2 ) 
  {$f_{i+1,j}^0$ $f_{i+1,j}^1$ $f_{i+1,j}^2$ $f_{i+1,j}^3$};
  \draw [ thick ] (coef.north west) -- (coef.north east);
  \draw [ thick ] (coef.south west) -- (coef.south east);
  \draw [ thick ] (coef.south west) -- (coef.north west);
  \draw [ thick ] (coef.south east) -- (coef.north east);
  \draw [ thin ] (coef.161) -- (coef.199);
  \draw [ thin ] (coef.90) -- (coef.270);
  \draw [ thin ] (coef.19) -- (coef.341);

  \draw [ thin ] ( #3, #4 ) -- (coef.south west);
  \draw [ thin ] ( #5, #6 ) -- (coef.south east);
}

\newcommand{\zerohcell}[6] {
  \node [ thin, scale = 1, draw ] (coef) at ( #1, #2 ) 
  {$0$ $0$ $0$ $0$};
  \draw [ thick ] (coef.north west) -- (coef.north east);
  \draw [ thick ] (coef.south west) -- (coef.south east);
  \draw [ thick ] (coef.south west) -- (coef.north west);
  \draw [ thick ] (coef.south east) -- (coef.north east);
  \draw [ thin ] (coef.140) -- (coef.220);
  \draw [ thin ] (coef.90) -- (coef.270);
  \draw [ thin ] (coef.40) -- (coef.320);

  \draw [ thin ] ( #3, #4 ) -- (coef.south west);
  \draw [ thin ] ( #5, #6 ) -- (coef.south east);
}

\begin{document}
\sloppy

\selectlanguage{english}

\title{Optimization of the Gyroaverage operator based on Hermite interpolation}

\author{F.\,Rozar}\address{Maison de la Simulation, USR 3441, CEA\,/\,CNRS\,/\,Inria\,/\,Univ.\,Paris-Sud\,/\,Univ.\,Versailles, 91191 Gif-sur-Yvette, FRANCE}\sameaddress{3}
\author{C.\,Steiner}\address{IRMA, Universit\'e de Strasbourg, France}
\author{G.\,Latu}\address{CEA, IRFM, F-13108 Saint-Paul-lez-Durance}
\author{M.\,Mehrenberger}\sameaddress{2}
\author{V.\,Grandgirard}\sameaddress{3}
\author{J.\,Bigot}\sameaddress{1}
\author{T.\,Cartier-Michaud}\sameaddress{3}
\author{J.\,Roman}\address{Inria, Bordeaux INP, CNRS, FR-33405 Talence}

%
%
\begin{abstract}
Gyrokinetic modeling is appropriate for describing Tokamak plasma
turbulence, and the gyroaverage operator is a cornerstone of this
approach. In a gyrokinetic code, the gyroaveraging scheme needs to be
accurate enough to avoid spoiling the data but also requires a low
computation cost because it is applied often  on
the main unknown, the $5D$ guiding-center distribution function, and
on the $3D$ electric potentials. In the present paper, we improve a
gyroaverage scheme based on Hermite interpolation used in the \gy
code. This initial implementation represents a too large fraction of the
total execution time. The gyroaverage operator has been reformulated
and is now expressed as a matrix-vector product and a cache-friendly algorithm has been
setup. Different techniques have been investigated to quicken the
computations by more than a factor two. Description of the algorithms
is given, together with an analysis of the achieved performance.
\end{abstract} 

%
%
%
\maketitle

\section{Introduction}

Gyrokinetic modeling is appropriate for describing Tokamak plasma
turbulence and the gyroaverage operator is a cornerstone of this
approach. This is the model used by \gy{}, a simulation code which 
is used to study the turbulence development in plasma fusion.
The gyroaverage operator $\mathcal{J}$ transforms the
so-called guiding-center distribution into the actual particle
distribution. It enables to take into account effects relative to 
the finite \textit{Larmor radius}, which is the radius of gyration of the gyro-center 
(motion which is faster than the turbulence
we are looking at). In the present paper, we improve the gyroaverage
scheme based on interpolation of the \gy code to speedup
calculations. For the current gyroaverage operator, 
the Larmor radius $\rho$ is assumed to be independent in space. 
This code considers a computational domain in five
dimensions (3D in space describing a torus geometry, 2D in
velocity)~\cite{Grandgirard_JOCP2006, Grandgirard_PPCF2007}. Time
evolution of the system consists in solving Vlasov equation
non-linearly coupled to a Poisson equation (electrostatic
approximation, quasi-neutrality is assumed). Routinely, physicists
perform large \gy simulations using from 1k to 16k cores on
supercomputers. This work aims at improving the
gyroaverage method based on Hermite interpolation which is too slow to
be used in production for the moment. In achieving this optimization,
we will allow the physicists to access to more accurate simulations
and then to better understand the physical processes that arise in
the simulations at finest scales.

The use of the Fourier transform reduces the gyroaveraging operation
by a multiplication in the Fourier space by the Bessel function. Good
approximations of the Bessel function have been proposed such as the
widely used \textit{Pad\'{e}} expansion. The Pad\'{e} approximation
enables to recover a good approximation for small radii (see
Fig.~\ref{fig:J0}) \,\cite{cms}. However, for a larger Larmor radius
and ordinary wave-numbers, the Pad\'{e} approximation truncates the
oscillations of the Bessel function, over-damps the small scales, and
then introduces bias in the simulation data. Also the use of Fourier
transform is not applicable in general geometry as we would
like~\cite{cms}; therefore it can not be employed in realistic tokamak
equilibrium. These two limitations are overcome using an interpolation
technique on the gyro-circles already introduced in~\cite{Steiner2014}
and which is the basis of our study.

In a gyrokinetic code, the gyroaveraging scheme needs to be accurate
enough to avoid spoiling the data but also requires low computation
cost because it is applied several times per time step on the main
unknown, the 5D guiding-center distribution function. The gyroaverage
is employed in \gy to compute the right-hand side of the
Poisson equation, the gyroaveraged electric potentials
that are used to get the advection field in the Vlasov solver and
several diagnostics that export physical quantities on mass
storage. One has to control the cost of this operator without
compromising the quality, but the numerical methods, the algorithms
and implementations play a major role here. By now, the new
implementation of the gyroaveraging is ready for production runs and some
numerical results are given.

The optimisation work presented here follows the previous
ones~\cite{Bigot_ESAIM2013, Latu_PPAM2012}. The work we have done on
the gyroaverage operator is presented in this article as follows. The
numerical method of the gyroaveraging using Hermite interpolation is
explained in Section~\ref{interpolation_method}. The operator has been
reformulated as a matrix-vector product in
Section~\ref{matricial_vision}.  Section~\ref{gyro_optim} details the
optimizations that quicken the computations by a factor two. Some
cache-friendly algorithms have been setup. Section~\ref{perf_results}
shows the performance results in the \gy code of the different
versions of the gyroaverage operator. Section~\ref{conclusion}
concludes and gives some hints to go forward in the optimization of
the gyroaverage operator.

\section{\label{interpolation_method}The gyroaverage operator based on interpolation method}

\subsection{\label{sec:Interpolation_method}Principle of the method}

\begin{figure}[t]
  \begin{center}
    \begin{tikzpicture}

      \foreach \i in {0, ..., 6 } {
        \draw ( \i, 0 ) arc ( 0 : 90 : \i );
      }

      \foreach \i in {0, ..., 6 } {
        \draw ( 0 + \i * 15 : 1) -- ( 0 + \i * 15 : 6 );
      }

      \cell{ 45 }{ 6 }
      \cell{ 60 }{ 5 }
      \cell{ 60 }{ 3 }
      \cell{ 15 }{ 3 }
      \cell{ 15 }{ 5 }

      \draw [ ->, >=latex, line width=0.8, color=gray!60 ] ( 15 : 6.2 ) -- ( 15 : 7.2 )
      node [ scale=1, above ] {{\color{black}$\vec{u_r}$}} ;

      \draw [ ->, >=latex, line width=0.8, color=gray!60 ] ( 15 : 6.2 ) 
      arc ( 15 : 30 : 6.2 )
      node [ scale=1, above ] {{\color{black}$\vec{u_\theta}$}} ;

      \draw [ ->, >=latex, color=gray!60 ] ( 0 : 1.2 ) arc ( 0 : 45 : 1.2 ) ;
      \draw [ color=gray!60 ] ( 0 : 0 ) -- ( 0 : 1.2 );
      \draw [ color=gray!60 ] ( 45 : 0 ) -- ( 45 : 1.2 );
      \node [ scale=0.8 ] at ( 7 : 2.0 ) {{\color{black}$\theta_3 = 45$\degre}};

      \draw [ ->, >=latex, color=gray!60 ] ( 0, -0.3 ) -- ( 4, -0.3 ) 
      node [ scale=0.8, midway, below ] {{\color{black}$r_3 = 4$}};
      \draw [ color=gray!60 ] ( 0, -0.1 ) -- ( 0, -0.6 );
      \draw [ color=gray!60 ] ( 4, -0.1 ) -- ( 4, -0.6 );

      \draw [ <->, >=latex, color=gray!60 ] ( 45 : 4 ) ++( 45 + 72 : 1.7 ) ++( 45 + 72 : -0.1 )
      -- ( 47 : 4.04 ) node [ scale=0.8, midway, above ] {{\color{black}$\rho$}} ;

      \draw [ ->, >=latex, color=gray!60 ] ( 45 : 4 ) ++( 0 : 0.6 ) arc ( 0 : 45 + 72 : 0.6 ) ;
      \node [ scale=0.8 ] at ( 48 : 4.70 ) {{\color{black}$\alpha_2$}} ;
      \draw [ color=gray!60 ] ( 45 : 4 ) -- ++( 0 : 1 ) ;

      \draw ( 45 : 4 ) circle (1.7);
      \draw ( 45 : 4 ) node [ scale = 1.4 ] {{\color{red}$\bullet$}};

      \draw ( 45 : 4 ) ++( 45 + 0 * 72 : 1.7 ) 
        node [ scale = 1 ] {{\color{blue}$\blacktriangle$}};
      \draw ( 45 : 4 ) ++( 45 + 1 * 72 : 1.7 ) 
        node [ scale = 1 ] {{\color{blue}$\blacktriangle$}};
      \draw ( 45 : 4 ) ++( 45 + 2 * 72 : 1.7 ) 
        node [ scale = 1 ] {{\color{blue}$\blacktriangle$}};
      \draw ( 45 : 4 ) ++( 45 + 3 * 72 : 1.7 ) 
        node [ scale = 1 ] {{\color{blue}$\blacktriangle$}};
      \draw ( 45 : 4 ) ++( 45 + 4 * 72 : 1.7 ) 
        node [ scale = 1 ] {{\color{blue}$\blacktriangle$}};

    \end{tikzpicture}

    \caption{
      Description of the gyroaverage based on an interpolation
      method. One has to compute the gyroaverage on the red point
      {\color{red}$\bullet$}. To do that, one estimates the average
      of the $f$ function at blue locations
      {\color{blue}$\blacktriangle$}. The $f$ values at blue points
      are obtained thanks to Hermite interpolation using $f$ which
      is known on the mesh points, here particularly at mesh points
      {\color{lightgreen}$\blacksquare$}.
    }
    \label{fig:gyroaverage_example}

  \end{center}
\end{figure}
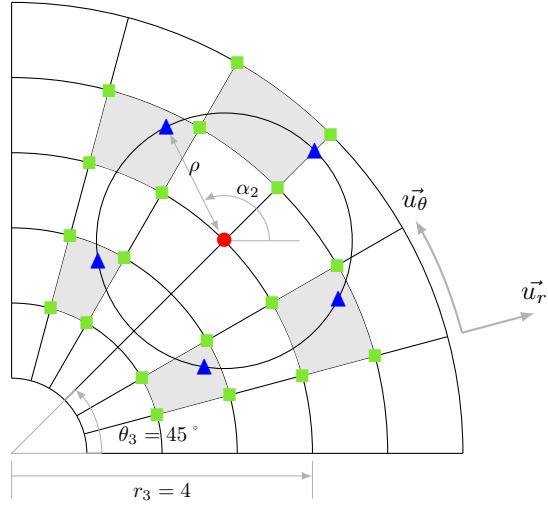

In this section, we describe the computation of the gyroaverage
operator in real space developed in \cite{Steiner2014}. This method
implies essentially interpolations over the Larmor circle. Let us
consider a function $f$ discretized over the $2D$ plane, on which the
gyroaverage operator will be applied in this study. We distribute uniformly $N$
quadrature points on the circle of integration and since the
quadrature points do not coincide with grid points, we introduce an
interpolation operator ${\cal P}$. Fig.~\ref{fig:gyroaverage_example}
gives an example of this approach with $N = 5$. We have evaluated two
schemes: Hermite interpolation and cubic spline interpolation.  The
gyroaverage is then obtained by the rectangle quadrature formula on
these points. More precisely, for a given point $(r_i, \theta_j)$ in a
polar coordinate system, the gyroaverage at this point is approximated
by
\begin{equation}
\mathcal{J}_\rho(f)_{r_i,\theta_j} \simeq \frac{1}{2\pi}\sum_{k=0}^{N-1} {\cal P}(f)(r_i \cos\theta_j+\rho \cos\alpha_k, r_i \sin\theta_j+\rho \sin\alpha_k) \Delta \alpha, 
\label{eq:general_gyro_formula}
\end{equation}

\noindent
where  $\alpha_k = \theta_j + k \Delta \alpha$ and $\Delta \alpha = 2\pi/N$. 

When points are outside the domain in the previous sum, we perform a
radial projection on the border of the domain:
\begin{itemize}
\item[-] if $r <r_ {\min}$ then ${\cal P}(f)(r, \theta)$ is replaced by ${\cal P}(f)(r_ {\min}, \theta),$
\item[-] if $r >r_ {\max}$ then ${\cal P}(f)(r, \theta)$ is replaced by ${\cal P}(f)(r_ {\max}, \theta).$
\end{itemize}
In the following, we describe the interpolation operator
$\mathcal{P}$ we used. In practice, the Hermite interpolation method is faster
than the cubic spline one \cite{Steiner2014}. So the 
optimizations done in this paper focus on the Hermite
interpolation method mainly, even if these optimizations can also be applied 
easily to the cubic spline approach.

\subsection{Hermite interpolation}

The Hermite interpolation method consists in a reconstruction of a
polynomial function of degree $3$ over a cell of the mesh (filled in
light gray on Fig.~\ref{fig:gyroaverage_example}) such that the value
and the first derivative of this polynomial are equal to those of the
input discretized function on the edges of the cell.

To do this, the values of the derivatives are reconstructed from the
nodal values of the function $f$ a finite difference
scheme of arbitrary order $d$.  Regardless of the $d$ value, the Hermite
polynomial remains of order $3$, which allows a flexible use.  It is
possible to improve the precision of the interpolation by increasing
the arbitrary order of the derivative reconstruction. In addition, unlike cubic splines interpolation, the
Hermite interpolation method is very local; it requires only very few 
points close to the target position of the
interpolation.

\subsubsection{Interpolation over a 1D mesh}
We detail  here the  Hermite interpolation operator  in the  uni-dimensional case
before taking into account the 2D polar coordinate system in the next subsection.
Let us consider a domain $\Omega =[a,b]\subset \mathbb{R}$ divided into $N$ cells:
$$
C_i= [x_{i},x_{i+1}], \quad \text{with } i \in \llbracket 0, N-1 \rrbracket .
$$
The mesh is assumed uniform, i.e. for any index $i$ the space step
$\Delta x$ verifies
$$
\Delta x = x_{i+1} -x_{i} = \frac{b-a}{N}.
$$ 
Let
$\alpha\in [0,1[$. The reconstruction of $f$ by Hermite interpolation
over the cell $C_i$ reads:

\begin{equation*}
  \begin{aligned}
    f(x_i + \alpha\Delta x) \ \approx \ & (2\alpha+1)(1-\alpha)^2 f(x_i) + 
                                          \alpha^2(3-2\alpha) f(x_{i+1})  + \\
                                        & \alpha(1-\alpha)^2 f'(x_i^+)   + 
                                          \alpha^2(\alpha-1) f'(x_{i+1}^-).
  \end{aligned}
\end{equation*}

The right and left derivatives are reconstructed by centered finite
differences of arbitrary order $d$ (see \cite{Li2005}).  In order to
compute the left (resp. right) derivative $f'(x_i^{-})$
(resp. $f'(x_i^{+})$) at the point $x_i$, we use the nodal values
$f(x_{i+k})$ around the point $x_i$ with a stencil from $k=r_d^{-}\leq
0$ to $k=s_d^{-}\geq 0$ (resp. $k=r_d^{+}\leq 0$ to $k=s_d^{+}\geq
0$). More precisely,
$$
f'(x_i^{-}) \approx \Psi^{-}_d(f(x_i)), \qquad f'(x_i^{+}) \approx \Psi^{+}_d(f(x_i))
$$
where the operators $\Psi^{\pm}_d : \mathbb{R}^{N+1} \rightarrow \mathbb{R}^{N+1}$ are:
$$ 
(\Psi^{-}_d(X))_i = \sum_{k=r_d^{-}}^{s_d^{-}}\omega_{k,d}^{-}
X_{i+k}, \qquad (\Psi^{+}_d(X))_i =
\sum_{k=r_d^{+}}^{s_d^{+}}\omega_{k,d}^{+} X_{i+k}
$$ 
with \\
$$
 \omega_{k,d}^{\pm} =
  \begin{dcases}
    \prod_{j=r_d^{\pm}, \ j\not=k,0}^{s_d^{\pm}} (-j) \bigg/  \prod_{j=r_d^{\pm}, \ j\not=k}^{s_d^{\pm}} (k-j) 
      & \text{if } k \neq 0 \\
    -\sum_{j=r_d^{\pm}, \ j\not= 0}^{s_d^{\pm}} \omega_{j,d}^{\pm}
      & \text{else.}
  \end{dcases}
$$

For an even reconstruction order $d=2p$, the stencil reads:
$$r_d^{-}=-p , \qquad s_d^{-}=p, \qquad r_d^{+}=-p+1 , \qquad
s_d^{+}=p+1$$ and for an odd order $d=2p+1$, we have:
$$r_d^{-}=r_d^{+}=-p , \qquad s_d^{-}=s_d^{+}=p+1.$$ We observe that
in the case of an odd (resp. even) order, the reconstruction is
uncentered (resp. centered) and the reconstructed derivative function
is $C^0$ (resp. $C^1$). 
In the numerical results presented in the
following, the default order will be $d=4$.

\subsubsection{\label{sec:2D_case}Interpolation over a 2D polar mesh}

We consider now an uniform polar mesh over the domain
$[r_{\min},r_{\max}]\times[0,2\pi]$ with $\Nr \times \Ntheta$ cells:
$$
C_{ij}=[r_i,r_{i+1}]\times[\theta_j,\theta_{j+1}], \qquad \text{with } i \in \llbracket 0, N_r-1\rrbracket, \, j \in \llbracket 0, N_{\theta}-1 \rrbracket
$$
where
\begin{eqnarray*} 
  r_i       = & r_{\min} + i\frac{r_{\max} - r_{\min}}{\Nr}, & i \in \llbracket 0, N_r \rrbracket \\
  \theta_j  = & j\frac{2\pi}{\Ntheta},                       & j \in \llbracket 0, N_{\theta}\rrbracket.
\end{eqnarray*}

Hermite interpolation over a polar mesh, which corresponds to a tensor
product, consists on a succession of one-dimensional Hermite
interpolations.  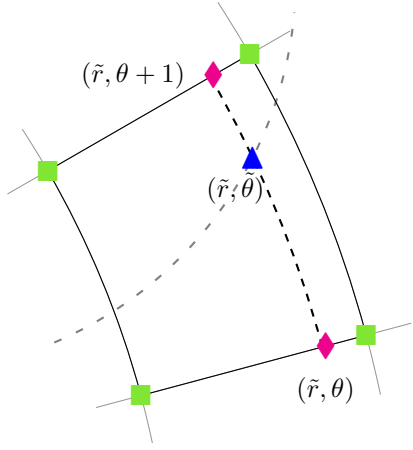
\begin{figure}[ht]
  \begin{center}

    \begin{tikzpicture}[scale=3.1]

      \draw [ color = gray!80 ] 
      ( 15 : 3.8 ) -- ( 15 : 5.2 );
      \draw [ color = gray!80 ] 
      ( 30 : 3.8 ) -- ( 30 : 5.2 );
      \draw [ color = gray!80 ] 
      ( 12 : 4 ) arc ( 12 : 33 : 4 );
      \draw [ color = gray!80 ] 
      ( 12 : 5 ) arc ( 12 : 33 : 5 );

      \draw [ loosely dashed, thick, color = gray!100 ] 
      ( 45 : 3.999 ) ++( 293 : 1.7 ) arc ( 293 : 355 : 1.7 );

      \cellclear{ 15 }{ 5 }

      \draw [ dashed, thick ] ( 15 : 4 ) ++( 15 : 0.805 ) arc ( 15 : 30 : 4.805 );

      \node [ scale=1.2 ] (bottom-pt) at ( 15 : 4 + 0.82 ) 
            { {\color{magenta}$\blacklozenge$} };
      \node [ scale=1, below=-0.1em of bottom-pt ] {$(\tilde{r}, \theta)$};

      \node [ scale=1.2 ] (top-pt) at ( 30 : 4 + 0.82 ) 
            { {\color{magenta}$\blacklozenge$} };
      \node [ scale=1, left=-0.1em of top-pt ] {$(\tilde{r}, \theta+1)$};

      \draw ( 45 : 3.999 ) ++( 288 + 45 : 1.7 ) 
      node [ scale=1.4 ] { {\color{blue}$\blacktriangle$} };
      \draw ( 45 : 3.999 ) ++( 288 + 45 - 5 : 1.7 ) 
      node [ scale=1, line width=1.5 ] { $(\tilde{r}, \tilde{\theta} )$ } ;

    \end{tikzpicture}

    \caption{
      Details on the interpolation steps for one point. The point
      {\color{blue}$\blacktriangle$} inside the cell represents the
      target value. It is obtained thanks to the Hermite interpolation
      between the intermediate points {\color{magenta}$\blacklozenge$}
      from the top and the bottom of the cell.
    }
    \label{fig:one_pt_interpolation}

  \end{center}
\end{figure}

Fig.~\ref{fig:one_pt_interpolation} highlights an intermediate
step. On the red points $(\tilde{r}, \theta)$ and $(\tilde{r},
\theta+1)$, the underlying discretized function $f$ is interpolated to
get its value and several derivatives. Then these values are used to
evaluate $f$ at the target point $f(\tilde{r},\tilde{\theta})$
generally inside a cell. The interpolation method is summed up by the
following algorithm:

\begin{itemize}
  \item[-] interpolation of the function $f(\cdot,\theta_j)$ over
    $[r_i,r_{i+1}]$ in order to evaluate $f(\tilde{r},\theta_j)$
  \item[-] interpolation of the function $f(\cdot,\theta_{j+1})$ over
    $[r_i,r_{i+1}]$ in order to evaluate $f(\tilde{r},\theta_{j+1})$
  \item[-] interpolation of the function $\partial_{\theta} f(\cdot,
    \theta_j^+)$ over $[r_i,r_{i+1}]$ in order to evaluate
    $\partial_{\theta} f(\tilde{r},\theta_j^+)$
  \item[-] interpolation of the function $\partial_{\theta} f(\cdot,
    \theta_{j+1}^-)$ over $[r_i,r_{i+1}]$ in order to evaluate
    $\partial_{\theta} f(\tilde{r},\theta_{j+1}^-)$
  \item[-] interpolation of the function $f(\tilde{r},\cdot)$ over
    $[\theta_j,\theta_{j+1}]$ by using the 4 previous evaluations to
    calculate $f(\tilde{r},\tilde{\theta})$.
\end{itemize}

\noindent
To achieve these 1D interpolations, we build as a first step the
partial derivatives at the cell interfaces:
\begin{eqnarray*}
(\partial_r f(r_i^+,\theta_j))_{i \in \llbracket 0, N_r \rrbracket} & \approx & \Psi^{+}_d(f(r_i,\theta_j)_{i \in \llbracket 0, N_r \rrbracket}) \qquad \forall j \in \llbracket 0, N_{\theta} \rrbracket\\
(\partial_r f(r_i^-,\theta_j))_{i \in \llbracket 0, N_r \rrbracket} & \approx & \Psi^{-}_d(f(r_i,\theta_j)_{i \in \llbracket 0, N_r \rrbracket}) \qquad \forall j \in \llbracket 0, N_{\theta} \rrbracket\\
(\partial_{\theta} f(r_i,\theta_j^+))_{j \in \llbracket 0, N_{\theta} \rrbracket} & \approx & \Psi^{+}_d(f(r_i,\theta_j)_{j \in \llbracket 0, N_{\theta} \rrbracket}) \qquad \forall i \in \llbracket 0, N_r \rrbracket\\
(\partial_{\theta} f(r_i,\theta_j^-))_{j \in \llbracket 0, N_{\theta} \rrbracket} & \approx & \Psi^{-}_d(f(r_i,\theta_j)_{j \in \llbracket 0, N_{\theta} \rrbracket}) \qquad \forall i \in \llbracket 0, N_r \rrbracket ,
\end{eqnarray*}
then the second derivatives: 
\begin{eqnarray*}
(\partial^2_{r,\theta} f(r_i^+,\theta_j^+))_{i \in \llbracket 0, N_r \rrbracket} & \approx & \Psi^{+}_d(\partial_{\theta} f(r_i,\theta_j^+)_{i \in \llbracket 0, N_r \rrbracket}) \qquad \forall j \in \llbracket 0, N_{\theta} \rrbracket\\
(\partial^2_{r,\theta} f(r_i^-,\theta_j^+))_{i \in \llbracket 0, N_r \rrbracket} & \approx & \Psi^{-}_d(\partial_{\theta} f(r_i,\theta_j^+)_{i \in \llbracket 0, N_r \rrbracket}) \qquad \forall j \in \llbracket 0, N_{\theta} \rrbracket\\
(\partial^2_{r,\theta} f(r_i^+,\theta_j^-))_{i \in \llbracket 0, N_r \rrbracket} & \approx & \Psi^{+}_d(\partial_{\theta} f(r_i,\theta_j^-)_{i \in \llbracket 0, N_r \rrbracket}) \qquad \forall j \in \llbracket 0, N_{\theta} \rrbracket\\
(\partial^2_{r,\theta} f(r_i^-,\theta_j^-))_{i \in \llbracket 0, N_r \rrbracket} & \approx & \Psi^{-}_d(\partial_{\theta} f(r_i,\theta_j^-)_{i \in \llbracket 0, N_r \rrbracket}) \qquad \forall j \in \llbracket 0, N_{\theta} \rrbracket.
\end{eqnarray*}

In the general case, for any node $(r_i,\theta_j)$, the following 9
values need to be recorded:
\begin{eqnarray*}
f(r_i,\theta_j), \quad \partial_r f(r_i^+,\theta_j), \quad \partial_r f(r_i^-,\theta_j), \quad \partial_{\theta} f(r_i,\theta_j^+), \quad \partial_{\theta} f(r_i,\theta_j^-)\\
\partial^2_{r,\theta} f(r_i^+,\theta_j^+),\quad \partial^2_{r,\theta} f(r_i^-,\theta_j^+), \quad \partial^2_{r,\theta} f(r_i^+,\theta_j^-), \quad \partial^2_{r,\theta} f(r_i^-,\theta_j^-).
\end{eqnarray*}
However, in the case of an even order $d$, as the reconstructed derivative
function is $C^1$, only the four following values are stored since $r^+=r^-$ and $\theta^+=\theta^-$:

\begin{equation}
f(r_i,\theta_j), \quad \partial_r f(r_i,\theta_j), \quad \partial_{\theta} f(r_i,\theta_j), \quad \partial^2_{r,\theta} f(r_i,\theta_j).
\label{eq:value-derivatives}
\end{equation}

\subsection{\label{sec:analytical_test_case}Analytical test case}

\begin{figure}[h!]
  \includegraphics[ width = 0.4\linewidth ]{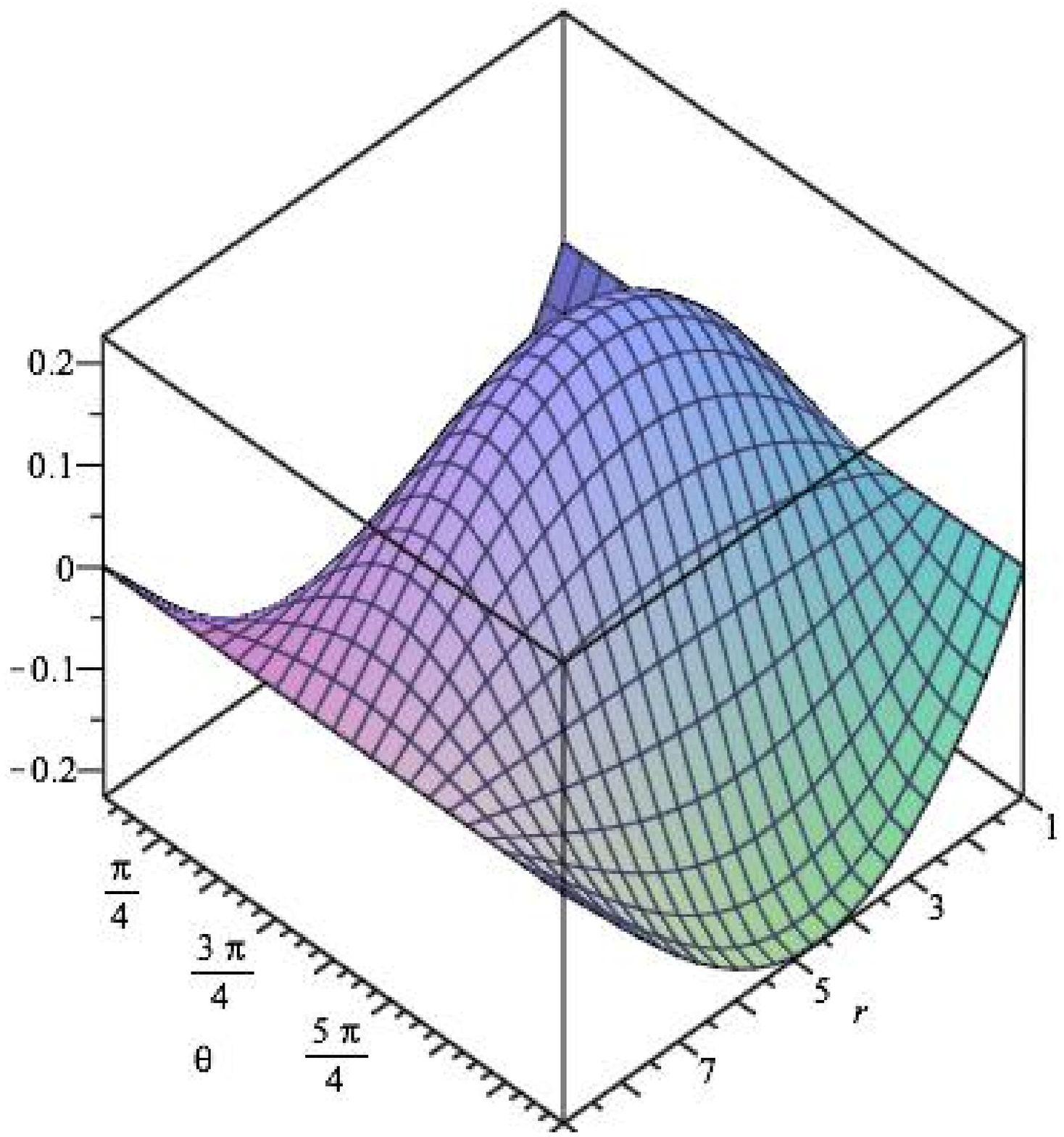}
  \includegraphics[ width = 0.4\linewidth ]{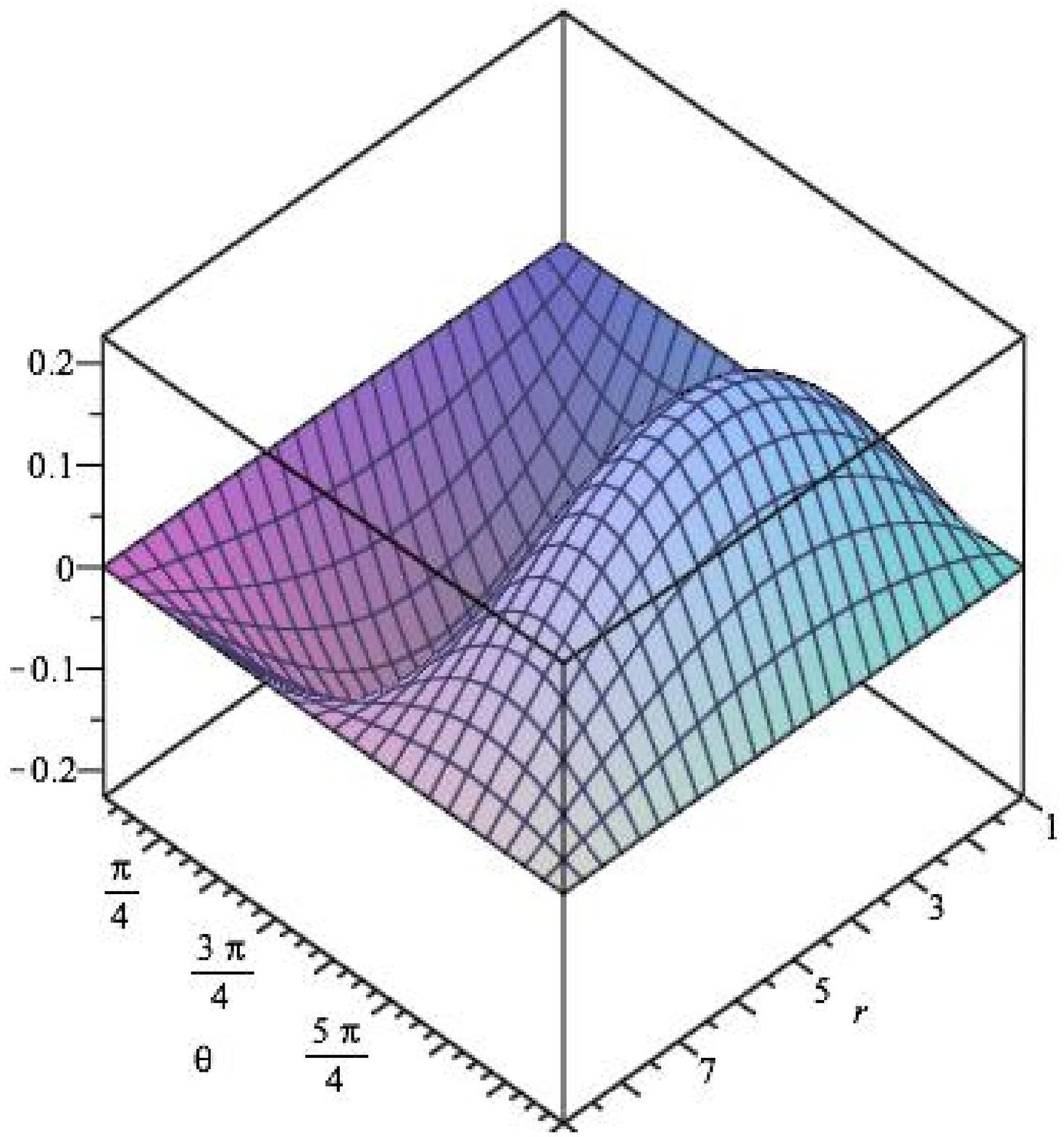}
  \caption{ 
    Real and imaginary parts of the test function $f_{1,1}$ with
    $r_{\min}=1$ and $r_{\max}=9$.
  }
  \label{fig:imre}
\end{figure}

To check and verify the gyroaveraging numerical solvers, one can use
analytical solutions. For that, we use the Fourier-Bessel functions
whose gyroaverage is analytically known \cite{Steiner2014}. More
precisely, for an integer $m\geq 0$, let $J_m$ and $Y_m$ be respectively the
Bessel functions of the first kind and of the second kind of order $m$
(see \cite{krehbessel}), we consider the following test function which
verifies the homogeneous Dirichlet conditions on $r_{\min}>0$ and
$r_{\max}$:
$$
f_{m,k}:(r,\theta)\in [r_{\min},r_{\max}]\times [0,2\pi] \mapsto 
\left( J_m(\gamma_{m,k}) Y_m\left( \gamma_{m,k}\frac{r}{r_{\max}} \right) - 
       Y_m(\gamma_{m,k}) J_m\left( \gamma_{m,k}\frac{r}{r_{\max}} \right) \right) e^{im\theta}
$$
where $\gamma_{m,k}$ is the $k^{th}$ zero of the function
$y\mapsto J_m(y)Y_m\left(y\frac{r_{\min}}{r_{\max}}\right)-Y_m(y)J_m\left(y\frac{r_{\min}}{r_{\max}}\right).$
Its gyroaverage reads:
\begin{eqnarray}
  \mathcal{J}_{\rho}(f_{m,k})(r_0,\theta_0) = 
  J_0\left(\gamma_{m,k}\frac{\rho}{r_{\max}}\right)f_{m,k}(r_0,\theta_0).
\label{eq:ana_sol}
\end{eqnarray}
The real and imaginary parts of the function $f_{1,1}$ is illustrated at Fig.~\ref{fig:imre}.

\begin{figure}[h!]
	\includegraphics [ width = 0.6\linewidth ] {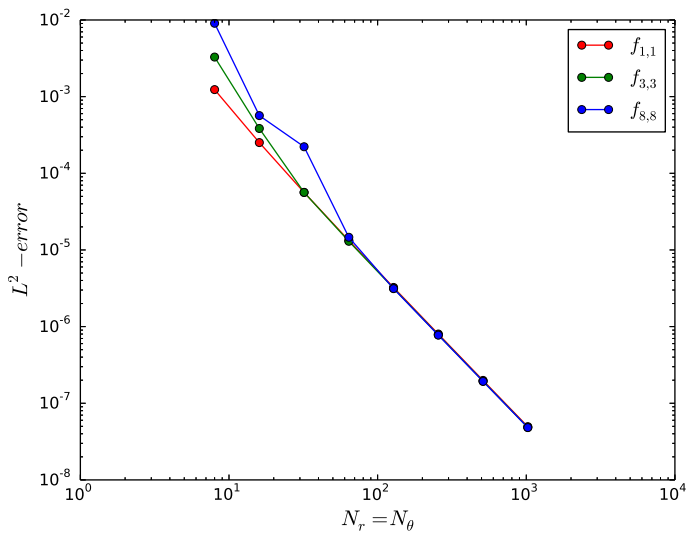}
	\caption{
		$L^2$-error for the gyroaverage of the real part of the functions
		$f_{1,1}, f_{3,3}$ and $f_{8,8}$ in function of $N_r=N_{\theta}$. The gyroaverage is computed using the method based on
		Hermite interpolation with $N=1024$ points on the Larmor
		circle. Parameters : $r_{\min}=1$, $r_{\min}=9$, $d=4$ and $\rho=1$.
	}
	\label{fig:order}
\end{figure}

We designed a unit test which is based on the analysis of the error --
the difference between analytical solution and the computed solution
(using $L^1$ or $L^2$ norm). Setting the gyroradius $\rho$, the number of points on the
Larmor circle $N$ and the interpolation order $d$, one can draw the
numerical behavior of the error and check the order~2 in space; in practice, we use the real part of these  functions.
The Fig.~\ref{fig:order} shows the $L^2 - error$ between the analytical gyroaverage and the numerically computed gyroaverage with the three functions $f_{1,1}$, $f_{3,3}$ and $f_{8,8}$ depending on the resolution of the  mesh. The higher the modes $(m,k)$ of a function are, the more this function oscillates, which induces larger approximation errors. This explains the observed differences between the curves on Fig.~\ref{fig:order} for the lowest values of $N_r,\ N_\theta$. Starting from $N_r = N_\theta \geq 10^{2}$, the three curves are not distinguishable any more, which means the resolution of the mesh is fine enough to catch the oscillations of the three functions and so to ensure a good accuracy of the gyroaverage computation.

\begin{figure}[h!]
  \includegraphics [ width = 0.6\linewidth ] {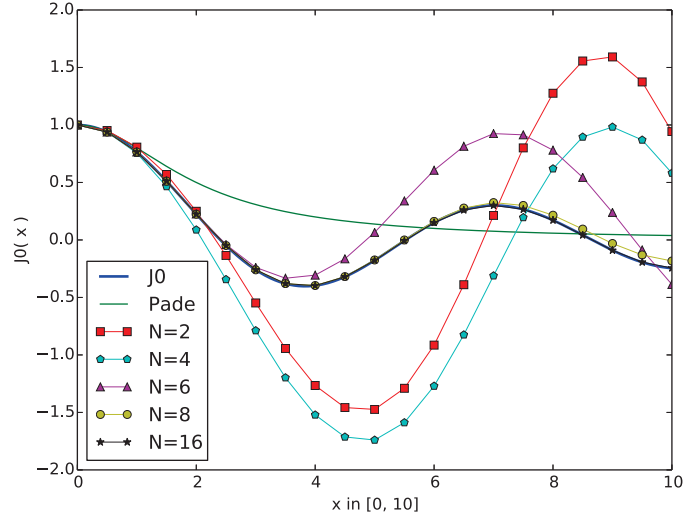}
  \caption{
    Reconstructions of the function $J_0$ using the function
    $f_{8,8}$ and its gyroaverage computed by the method based on
    Hermite interpolation with $N=2,4,6,8$ or $16$ points on the Larmor
    circle. These reconstructions are functions of $x=\gamma_{8,8}\frac{\rho}{r_{\max}}.$
  }
  \label{fig:J0}
\end{figure}

Another verification consists, for a function $f_{m,k}$ and a given
point $(r_0,\theta_0)$, in calculating the ratio
$$
\frac{\mathcal{J}_{\rho}(f_{m,k})(r_0,\theta_0)}
     {f_{m,k}(r_0,\theta_0)} \approx 
     J_0\left(\gamma_{m,k}\frac{\rho}{r_{\max}}\right)
$$
which can be seen as an approximation of the Bessel function $J_0$
thanks to Eq.~\eqref{eq:ana_sol}. Fig.~\ref{fig:J0} illustrates the
reconstruction of $J_0$ using the function $f_{8,8}$ with $r_{\min}=1$, $r_{\max}=9$, $N_r=N_{\theta}=1024$, $(r_0,\theta_0)$ has coordinates $(470,470)$ in the mesh, $\gamma_{8,8}\approx 36.0$ and $0\leq \rho \leq \frac{10 r_{\max}}{\gamma_{8,8}}\approx 2.5$. Since the function $f_{8,8}$ has the highest mode values among the three considred functions, it is the most difficult function to handle by the gyroaverage operator and so it is relevant to study the error depending on the number of interpolation points.
The larger the number $N$ is, the better the Hermite curves fit the analytical solution $J_0$, whereas the Pad\'e approximation appears as rough starting from $x=2$.

These unit tests allow us to study the precision of the gyroaverage
operator depending on the input parameters but it also represents 
reference cases. The validity of any new implementation was
systematically checked by comparing their results with these unit
tests against those of the reference implementation. During the
development process, this check warns us early if something goes
wrong, it facilitates the fixing of bugs.

\section{\label{matricial_vision}Matrix representation for gyroaveraging}

\subsection{\label{sec:matrix_desc}Matrix description of the gyroaverage operator}
The previous section described the gyroaverage computed at a given
grid point thanks to Eq.~\eqref{eq:general_gyro_formula}. In the case of the even
order, this formula boils down to a linear combination of $4$ values
at the nodal points of the polar mesh: the value, the derivative of
first order in the radial and in the poloidal directions, and the
cross-derivative of second order (see
relation~\eqref{eq:value-derivatives}). These values are denoted
\rhsval{} in the following.

In order to figure out how to compute the gyroaverage of all the
points in the poloidal plane, a matrix representation of the operator
is useful. Using this formulation, we will highlight the properties of
the operator used for optimization issues. The gyroaverage based on
Hermite interpolation can be written as follows:
\begin{gather*}
  M_{final} = M_{coef} \times M_{fval} \\
  \mathrm{with}
  \begin{cases} 
    M_{final} & \in \mathcal{M}_{\Nr+1      , \Ntheta    } (\mathbb{R}), \\
    M_{coef}  & \in \mathcal{M}_{\Nr +1       , 4(\Nr+1)\Ntheta} (\mathbb{R}), \\
    M_{fval}  & \in \mathcal{M}_{4(\Nr+1)\Ntheta, \Ntheta    } (\mathbb{R}).
  \end{cases} 
\end{gather*}
\noindent
The matrix $M_{final}$ is the result of the gyroaverage
applied to all the points of a poloidal cut. Every point on this
matrix verifies the relation:

$$
  M_{final}(i,\,j) = \mathcal{J}_\rho(f)_{r_i,\theta_j} 
$$
\[\mathrm{with}\left\{
  \begin{array}{rl l}
    r_i      = & r_{\min} + i \frac{r_{\max} - r_{\min}}{\Nr}, & i \in \llbracket 0, \Nr \rrbracket, \\
    \theta_j = & j \frac{2\pi}{\Ntheta}, & j \in \llbracket 0, \Ntheta - 1 \rrbracket.
  \end{array}
\right.
\]
\vspace{1em}

The last point in the $\theta$ direction is excluded as it shares the same
position as the first point in this periodical direction. Each row
$R_{i}$ of the matrix $M_{coef}$ contains the contribution
coefficients associated to the input plan at a given index $i$ in
the radial direction. The matrix $M_{fval}$ contains in each
column $C_j$ the required values for Hermite
interpolation. The factor $4$ which appears in the size of the
matrices $M_{coef}$ and $M_{fval}$ is due to the
number of required values by the interpolation method over a $2D$
plane -- $4$ per grid point in the Hermite case.

With this representation, the gyroaverage computed at one grid point can be 
expressed as the following scalar dot product:
\begin{equation}
  \mathcal{J}_\rho(f)_{r_i,\theta_j} = 
  \mathnormal{R_{i}}(M_{coef}) \cdot \mathnormal{C_{j}}(M_{fval}).
  \label{eq:dot_product}
\end{equation}

\subsection{\label{sec:initial_implementation}Initial implementation of the gyroaverage operator}
For convenience, we will use the matrix-like notations introduced
above to describe the initial implementation used in
\cite{Steiner2014}. This implementation does not benefit of the clear
identification of the manipulated objects. In the initial implementation of 
the algorithm, the gyroaverage operator was not seen as a scalar dot product, 
and so some contribution was not factorized in the most efficient way.

Accordingly to the matrix representation of the gyroaverage, the
two main properties of the matrix $M_{coef}$ are
\textit{(i)} the sparsity (section~\ref{sec:mcoef_pattern}) and
\textit{(ii)} and the fact that it remains unchanged during the whole
simulation.  This matrix is computed once for all in the
initialization step, whereas the matrix $M_{fval}$ is
computed at each gyroaveraging because the values it contains are determined by
the considered input data.

The implementation is composed of two steps: the first at the
initialization stage described by Algo.~\ref{algo:initialisation} and
the second at each call to \texttt{gyro\_compute} during the
simulation run described by Algo.~\ref{algo:gyro_compute}.

\begin{algorithm}
  \SetKwData{Pt}{pt}
  \SetKwData{Cell}{cell}
  \SetKwInOut{Input}{input}
  \SetKwInOut{Output}{output}
  \Input{ $\Nr$, $\Ntheta$, $\rho$, $N$ }
  \Output{
    A representation of $M_{coef}$ -- the weights and the indexes of their corresponding \rhsval{} for each $i$.
  }

  \BlankLine

  \For{ $i$ $\leftarrow 0$ \KwTo $\Nr$ }{
    \For{ $k \leftarrow 0$ \KwTo $N-1$ }{
      Compute the $k^{th}$ position of the quadrature point \Pt over the Larmor circle of center ($i$, $j=0$)\;
      Compute/store the indexes of the corner points of the cell \Cell containing \Pt\;
      Compute/store the weights associated to the contribution of each corner point of \Cell into a representation of $M_{coef}$\;
    }
  }

  \caption{Computation of a representation of $M_{coef}$ (initialization step).}
  \label{algo:initialisation}
\end{algorithm}
\vspace*{-0.5em}

\begin{algorithm}
  \SetKwData{ir}{$i$}
  \SetKwData{Pt}{pt ($i$, $j$)}
  \SetKwData{IP}{input\_plane}
  \SetKwInOut{Input}{input}
  \SetKwInOut{Output}{output}
  \Input { The $f$ values in the poloidal plane \IP }
  \Output{ The gyroaveraged poloidal plane: gyroaveraged values at each grid point of the poloidal plane}

  \BlankLine

  \For{ each grid point \Pt of \IP }{
    Compute/store the first order derivative of $f$ in each dir. and the cross-derivative at \Pt into a representation of $\mathnormal{C_0}(M_{fval})$\;
  }

  \For{ each point \Pt of \IP }{
    Compute the gyroaveraged value at \Pt -- do the sparse dot product $\mathnormal{R_{i}}(M_{coef}) \times \mathnormal{C_{j}}(M_{fval})$\;
    Store the gyroaveraged value into the \IP at \Pt\;
  }

  \caption{
    Computation of the gyroaverage of $f$ over a poloidal plane taken as input.
  }
  \label{algo:gyro_compute}
\end{algorithm}
\vspace*{-1em}

\subsection{\label{sec:mcoef_pattern}Pattern of the matrix $M_{coef}$}

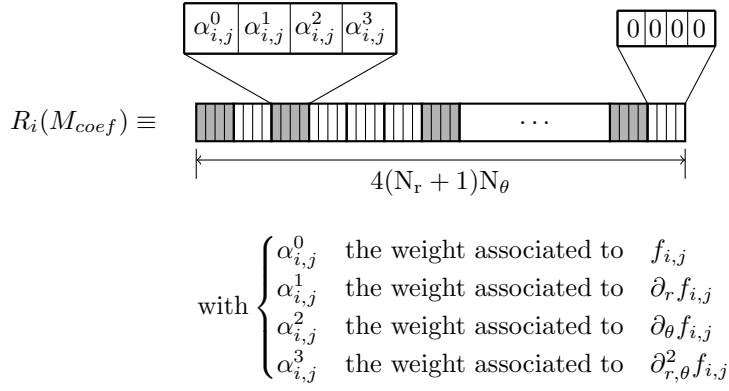
\begin{figure}[h!]
  \begin{center}
    \begin{tikzpicture}

      \filledhcell{ 0 + 0 * 0.5 }{ 0 }{ 0.5 }{ 0.5 }
      \filledhcell{ 0 + 2 * 0.5 }{ 0 }{ 0.5 }{ 0.5 }
      \filledhcell{ 0 + 6 * 0.5 }{ 0 }{ 0.5 }{ 0.5 }
      \filledhcell{ 0 + 11 * 0.5 }{ 0 }{ 0.5 }{ 0.5 }

      \foreach \i in {0,...,6} {
        \hcell{ 0 + \i * 0.5 }{ 0 }{ 0.5 }{ 0.5 }
      }
      \dothcell{ 0 + 7 * 0.5 }{ 0 }{ 2 }{ 0.5 }
      \foreach \i in {11,...,12} {
        \hcell{ 0 + \i * 0.5 }{ 0 }{ 0.5 }{ 0.5 }
      }

      \alphahcell{ 0 + 2 * 0.5 + 0.25 }{ 1.5 }{ 2 * 0.5 }{ 0.5 }{ 3 * 0.5 }{ 0.5 }
      \zerohcell{ 0 + 12 * 0.5 + 0.25 }{ 1.5 }{ 12 * 0.5 }{ 0.5 }{ 13 * 0.5 }{ 0.5 }

      \draw [ thin ] (  0      , -0.1 ) -- (        0, -0.5 );
      \draw [ thin ] ( 13 * 0.5, -0.1 ) -- ( 13 * 0.5, -0.5 );
      \draw [ thin, <->, =>latex ] ( 0, -0.25 ) -- ( 13 * 0.5, -0.25 )
      node [ midway, below ] {$ 4 (\Nr+1) \Ntheta $};

      \node at ( -1.5, 0.25 ) { $\mathnormal{R_{i}}(M_{coef}) \equiv$ } ;

      \node at ( 3.75, -2 ) (box){
      \begin{minipage}{0.50\textwidth}
        \begin{gather*}
          \mathrm{with}
          \begin{cases} 
            \alpha_{i,j}^0\quad \text{the weight associated to}\quad f_{i,j}\\
            \alpha_{i,j}^1\quad \text{the weight associated to}\quad \partial_{r} f_{i,j} \\
            \alpha_{i,j}^2\quad \text{the weight associated to}\quad \partial_{\theta} f_{i,j} \\ 
            \alpha_{i,j}^3\quad \text{the weight associated to}\quad \partial^2_{r, \theta} f_{i,j}
          \end{cases}
        \end{gather*}  
      \end{minipage}
      };
  
    \end{tikzpicture}

    \caption{
      Sketch of the sparsity of a row vector
      $\mathnormal{R_{i}}(M_{coef})$. On this example,
      only the cells filled in grey contain non-zero values
      $\alpha_{i,j}^*$.
    }
    \label{fig:sparse_vector}

  \end{center}
\end{figure}

To figure out the sparsity of the matrix $M_{coef}$, take a
look at Fig.~\ref{fig:gyroaverage_example}
(p.~\pageref{fig:gyroaverage_example}). The $i$-th line of the
matrix $M_{coef}$ contains the coefficient contribution of
every position of the input poloidal plane for the gyroaverage at a
point $(i,j\!=\!*)$. As we can see on
Fig.~\ref{fig:gyroaverage_example}, to compute the gyroaverage of the
point in the center of the circle ({\color{red}$\bullet$}), only a few
points of the plane contribute
({\color{lightgreen}$\blacksquare$}). As a consequence, the weight
associated to a point of the plan is non-zero for this gyroaverage
only if it is used to get the value of an interpolate point
({\color{blue}$\blacktriangle$}). This is the case for any mesh
point. This leads us to conclude that the matrix $M_{coef}$
is sparse. The Fig.~\ref{fig:sparse_vector} illustrates the generic
appearance of the $i$-th line vector of $M_{coef}$.

The value of the contribution coefficients is invariant with respect
to the poloidal index $j$. This is due to the fact that for $2$
separate mesh points ($i$,~$j_{1}$) and
($i$,~$j_{2}$), the polar coordinates of the interpolation
points involved in their respective gyroaverage remain the same in
local coordinate system along poloidal direction
($\vec{u_r}$,~$\vec{u_\theta}$). The value and the distribution of the
$\alpha_{i,j}^*$ coefficients depend only of the radial index
$i$.

The number of non-zero values of $2$ separated row vectors may be
different. The pattern of the cells which contribute to the
computation of the gyroaverage depends on the radial index $i$ of
the target point. Generally, the target points close to the center of
the poloidal mesh (i.e. $i$ is low) involve a larger number of cells compared to the
point near to the external boundary of the plane.

\subsection{\label{pattern_matrix_fval}Pattern of the matrix $M_{fval}$}

\begin{figure}[h!]
  \begin{center}
    \begin{tikzpicture}

      \foreach \i in {0,...,6} {
        \filledhcell{ 0 + \i * 0.5 }{ 0 }{ 0.5 }{ 0.5 }
        \hcell{ 0 + \i * 0.5 }{ 0 }{ 0.5 }{ 0.5 }
      }
      \dothcell{ 0 + 7 * 0.5 }{ 0 }{ 2 }{ 0.5 }
      \foreach \i in {11,...,12} {
        \filledhcell{ 0 + \i * 0.5 }{ 0 }{ 0.5 }{ 0.5 }
        \hcell{ 0 + \i * 0.5 }{ 0 }{ 0.5 }{ 0.5 }
      }

      \fvalhcell    { 0 - 0 * 0.5 }{ 1.5 }{ 3 * 0.5 }{ 0.5 }{ 4 * 0.5 }{ 0.5 }
      \nextfvalhcell{ 0 + 9 * 0.5 }{ 1.5 }{ 4 * 0.5 }{ 0.5 }{ 5 * 0.5 }{ 0.5 }

      \draw [ thin ] (  0      , -0.1 ) -- (        0, -0.5 );
      \draw [ thin ] ( 13 * 0.5, -0.1 ) -- ( 13 * 0.5, -0.5 );
      \draw [ thin, <->, =>latex ] ( 0, -0.25 ) -- ( 13 * 0.5, -0.25 )
      node [ midway, below ] {$ 4 (\Nr+1) \Ntheta $};

      \node at ( -1.5, 0.25 ) { $\mathnormal{C_{j}}(M_{fval}) \equiv$ } ;

      \node at ( 3.75, -2 ) (box){
      \begin{minipage}{0.50\textwidth}
        \begin{gather*}
          \mathrm{with}
            \begin{cases} 
              f_{i,j}^0 = f_{i,j} \\ 
              f_{i,j}^1 = \partial_{r} f_{i,j} \\ 
              f_{i,j}^2 = \partial_{\theta} f_{i,j} \\
              f_{i,j}^3 = \partial^2_{r,\theta} f_{i,j}
            \end{cases}
        \end{gather*}  
      \end{minipage}
      };
  
    \end{tikzpicture}

    \caption{
      Sketch of the layout of a column vectors of
      $M_{fval}$.
    }
    \label{fig:value_vector}

  \end{center}
\end{figure}
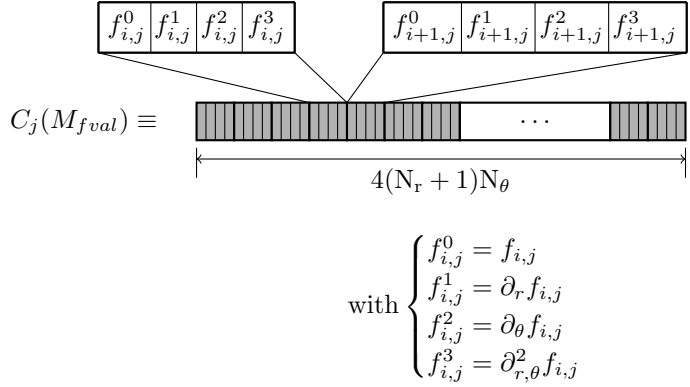

The matrix $M_{fval}$ stores the needed values for the
Hermite interpolation, namely the nodal value and their
derivatives. The layout of a column of this matrix is illustrated
on Fig.~\ref{fig:value_vector}.

The length of any vector
$\mathnormal{C_{j}}(M_{fval})$ is $4 (\Nr+1) \Ntheta$
and contains only non-zero values. In the current matrix
representation (see Fig.~\ref{fig:mfval_pattern}), one can notice that
$M_{fval}$ has a repetitive pattern. The distribution of the
values in a vector $\mathnormal{C_{j}}(M_{fval})$
depends on its poloidal index $j$. The values of the vector
$\mathnormal{C_{j+1}}(M_{fval})$ can be obtained by
a circular permutation of $4(\Nr+1)$ positions of the values of
$\mathnormal{C_{j}}(M_{fval})$. The pattern of
$M_{fval}$ is outlined on
Fig.~\ref{fig:mfval_pattern}. From the $1^{st}$ column vector, the
other ones can be deducted thanks to the following relation:
\begin{equation}
\mathnormal{C_{j \neq 0}}(M_{fval}) [ i ] = 
  \mathnormal{C_0}(M_{fval}) [ i - j \times 4(\Nr+1) ]
  \label{eq:fval_column_relation}
\end{equation}

\tikzset{
    node style value/.style={ 
      inner ysep=-1pt, 
      inner xsep=10pt, 
      minimum width = 45pt, 
      outer sep=2},
    arrow style shift/.style={draw, circle, midway,
    fill=white, opacity=0.9, inner sep=1, outer sep=0},
    arrow style ar/.style={ ->, red, opacity=0.5 },
}

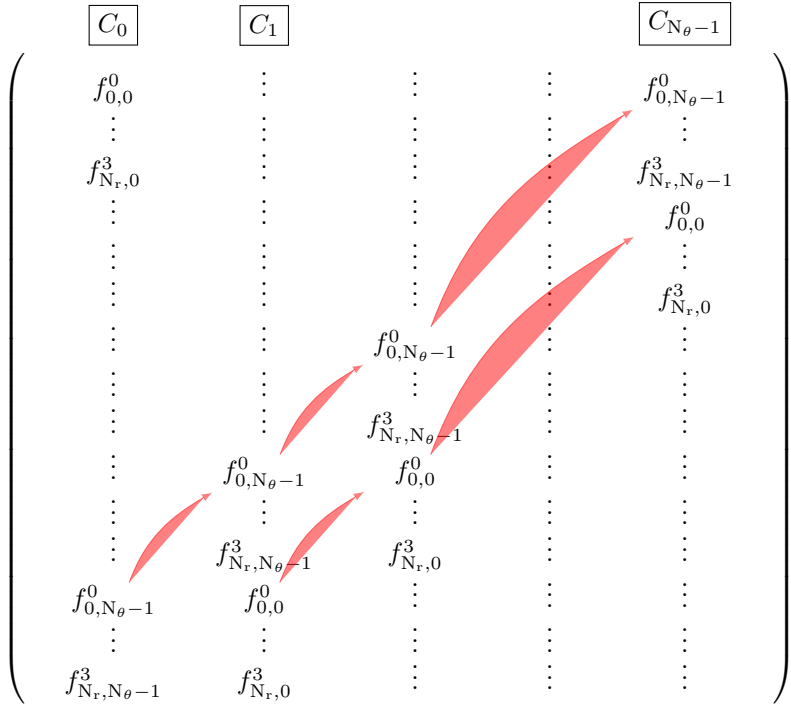
\begin{figure}[h]
  \begin{center}
    \begin{tikzpicture}[>=latex]
    
    \matrix (A) [matrix of math nodes,%
                 nodes = {node style value},%
                 left delimiter  = (,%
                 right delimiter = )]
    {%
      f_{0,0}^0           & \vdots              & \vdots              & \vdots   & f_{0,\Ntheta-1}^0   \\
      \vdots              & \vdots              & \vdots              & \vdots   & \vdots              \\
      f_{\Nr,0}^3         & \vdots              & \vdots              & \vdots   & f_{\Nr,\Ntheta-1}^3 \\
    		   	  	       	             
      \vdots              & \vdots              & \vdots              & \vdots   & f_{0,0}^0           \\
      \vdots              & \vdots              & \vdots              & \vdots   & \vdots              \\
      \vdots              & \vdots              & \vdots              & \vdots   & f_{\Nr,0}^3         \\
    			  		       							      
      \vdots              & \vdots              & f_{0,\Ntheta-1}^0   & \vdots   & \vdots              \\
      \vdots              & \vdots              & \vdots              & \vdots   & \vdots              \\
      \vdots              & \vdots              & f_{\Nr,\Ntheta-1}^3 & \vdots   & \vdots              \\
    		   	  		           						      
      \vdots              & f_{0,\Ntheta-1}^0   & f_{0,0}^0           & \vdots   & \vdots              \\
      \vdots              & \vdots              & \vdots              & \vdots   & \vdots              \\
      \vdots              & f_{\Nr,\Ntheta-1}^3 & f_{\Nr,0}^3         & \vdots   & \vdots              \\
                                                                   				      
      f_{0,\Ntheta-1}^0   & f_{0,0}^0           & \vdots              & \vdots   & \vdots              \\
      \vdots              & \vdots              & \vdots              & \vdots   & \vdots              \\
      f_{\Nr,\Ntheta-1}^3 & f_{\Nr,0}^3         & \vdots              & \vdots   & \vdots              \\
    };
    
    \draw[ arrow style ar ] (A-10-2.50) to[ out=70, in=210 ] (A-7-3.200);
    \draw[ arrow style ar ] (A-13-1.50) to[ out=70, in=210 ] (A-10-2.200);
    \draw[ arrow style ar ] (A-13-2.50) to[ out=70, in=210 ] (A-10-3.200);

    \draw[ arrow style ar ] (A-7-3.50) to[ out=70, in=210 ] (A-1-5.200);
    \draw[ arrow style ar ] (A-10-3.50) to[ out=70, in=210 ] (A-4-5.200);

    \node [ draw, above=10pt ] at (A-1-1.north) 
        { $C_0$ };
    \node [ draw, above=3pt ] at (A-1-2.north) 
        { $C_1$ };
    \node [ draw, above=10pt ] at (A-1-5.north) 
        { $C_{\Ntheta-1}$ };
    
    \end{tikzpicture}

    \caption{
      Illustration of the value position shift between the columns of $M_{fval}$.
    }
    \label{fig:mfval_pattern}

  \end{center}
\end{figure} 

The matrix representation of the gyroaverage operator provided in this
Section gives us a useful description. Section~\ref{gyro_optim}
details the different optimizations we have made to quicken the
initial computation of the gyroaverage thanks to this matrix representation.

\section{\label{gyro_optim}Optimize and speedup gyroaveraging}

The gyroaverage operator is decomposed into two stages, as already
mentioned in section~\ref{sec:initial_implementation}.
Algo.~\ref{algo:initialisation} is computed during the initialization
step and only once, so its execution time is taken as a payload for
the simulation and is not the goal of the optimizations presented
here. Algo.~\ref{algo:gyro_compute} is called many times during an
iteration of \gy and is the main kernel of the gyroaverage
operator. The optimizations done focus on the reduction of the
execution time of this algorithm. To achieve the optimization of the
gyroaverage operator, we paid attention to data structures which
represent the matrices involved in the computation. In a second step,
issues relative to cache memory levels are tackled.

In this Section, we describe several implementations. The first
one that is  presented in~\ref{sec:compact_vector} must be seen as
the starting point for the optimizations introduced in~\ref{sec:blocking}
and~\ref{sec:layout}.

\subsection{\label{sec:compact_vector}Compact vector optimization}

Obviously, the choice of the data structures for the matrix
$M_{coef}$ and $M_{fval}$ has great impact on the
performance. As said in~\ref{sec:mcoef_pattern}, the matrix
$M_{coef}$ is sparse. To benefit of this property, only
non-zero values of $M_{coef}$ are taken into account. By
avoiding the storage of null values, the size of vectors involved in
Eq.~\eqref{eq:dot_product} is reduced, but it requires the use of
dedicated data structures to achieve the computation of
Algo.~\ref{algo:gyro_compute}.

The main contribution of this optimization is the choice of the data
structure that implements the sparse matrix version of the
gyroaverage.

During the development process, different data structures were tested.
The data structure which represents the matrix $M_{coef}$
is less challenging than the representation of the matrix
$M_{fval}$, since $M_{fval}$ has a visible impact on
performances. Among the different representations of 
$M_{fval}$, two of them are detailed and compared below.

\subsubsection{The representation of the matrix $M_{coef}$}

The goal of the computation of Algo.~\ref{algo:initialisation} is
to compute the representation of the matrix $M_{coef}$. As
said previously, it is done during the initialization step and only
once. To benefit of the sparsity of this matrix $M_{coef}$,
a dedicated data structure \texttt{contribution\_vector} has been
created to handle it. This structure is defined as follows:
\begin{listing}{1}
  type :: gyro_vector
     integer, dimension(:), pointer :: ind
     real(8), dimension(:), pointer :: val
  end type gyro_vector

  type(gyro_vector), dimension(:), pointer :: contribution_vector
\end{listing}

The variable \texttt{contribution\_vector} is an array of length
$(\Nr+1)$ and of type \texttt{gyro\_vector} which is a structure as
defined above. Each cell of \texttt{contribution\_vector} represents a
row vector of the matrix $M_{coef}$. For a given radial
index $i$, the indexes of the underlying mesh points which
contribute to the gyroaverage of the point ($i$,~$j\!=\!0$)
are computed and saved
in \texttt{contribution\_vector(}$i$\texttt{)\%ind}. Its length is
denoted by \texttt{nb\_contribution\_pt(}$i$\texttt{)} and may be
different for each radial index (see
section~\ref{sec:mcoef_pattern}). These indexes are used to build the
variable which represents the matrix $M_{fval}$ which will
be detailed below. The weights associated to these positions are
computed and stored
in \texttt{contribution\_vector(}$i$\texttt{)\%val}.

\subsubsection{\label{first_representation}First representation of the matrix $M_{fval}$}

\tikzset{ 
    arrow style ar/.style={ <->, >=latex, color=gray!80, thick },
    arrow style cote/.style={ <->, >=latex, very thin },
    node style textbelow/.style={ below, sloped, scale=0.7 },
    node style textabove/.style={ above, sloped, scale=0.9 },
    node style multi/.style={ draw, circle, fill=white, midway, inner sep=1 },
}

\begin{figure}[t]
  \begin{center}
    \begin{tikzpicture}[ scale=0.8 ]
      \def\width{ 1 }

      \foreach \i in {0,1,...,2} {
        \hcell { \i * \width + 0 }{ 0 }{ \width }{ 0.5 }
      }
      \dothcell{ 3 * \width + 0 }{ 0 }{ 2 * \width }{ 0.5 }
      \hcell   { 5 * \width + 0 }{ 0 }{ \width }{ 0.5 }

      \foreach \j in {0,1,...,2} {
        \vcell   { \j * 0.5 + 6.5 * \width }{ 0 * \width + 1 }{ 0.5 }{ \width }
        \dotvcell{ \j * 0.5 + 6.5 * \width }{ 1 * \width + 1 }{ 0.5 }{ 2 * \width }
        \foreach \i in {3,...,5} {
          \vcell{ \j * 0.5 + 6.5 * \width }{ \i * \width + 1 }{ 0.5 }{ \width }
        }
      }

      \draw [ thick ] ( 3 * 0.5 + 6.5 * \width, 7 * \width ) -- ++( 3 * 0.5, 0 );
      \draw [ thick ] ( 3 * 0.5 + 6.5 * \width, 1 * \width ) -- ++( 3 * 0.5, 0 );
      \path (  4 * 0.5 + 6.5 * \width + 0.2, 3 * \width + 1 ) node [ scale=2 ]
      { $\ldots$ };

      \foreach \j in {6,...,6} {
        \vcell   { \j * 0.5 + 6.5 * \width }{ 0 * \width + 1 }{ 0.5 }{ \width }
        \dotvcell{ \j * 0.5 + 6.5 * \width }{ 1 * \width + 1 }{ 0.5 }{ 2 * \width }
        \foreach \i in {3,...,5} {
          \vcell{ \j * 0.5 + 6.5 * \width }{ \i * \width + 1 }{ 0.5 }{ \width }
        }
      }

      \draw ( 0, 0 - 0.1 ) -- ( 0, -0.5 );
      \draw ( 6, 0 - 0.1 ) -- ( 6, -0.5 );
      \draw[ arrow style cote ] ( 0 + 0.1, -0.25 ) to 
      node [ node style textbelow ] {\texttt{nb\_contribution\_pt(}$i$\texttt{)} $\times 4$}
      ( 6 - 0.1, -0.25 );

      \draw ( 7 * 0.5 + 6.5 * \width + 0.1, 1 ) -- ( 7 * 0.5 + 6.5 * \width + 0.5, 1 );
      \draw ( 7 * 0.5 + 6.5 * \width + 0.1, 7 ) -- ( 7 * 0.5 + 6.5 * \width + 0.5, 7 );
      \draw[ arrow style cote ] ( 7 * 0.5 + 6.5 * \width + 0.25, 1 + 0.1 ) to 
      node [ node style textbelow ] {\texttt{nb\_contribution\_pt(}$i$\texttt{)} $\times 4$}
      ( 7 * 0.5 + 6.5 * \width + 0.25, 7 - 0.1 );

      \foreach \i in {0,...,2} {
        \draw[ arrow style ar ] ( \i * \width + 0.5, 0.5 + 0.1 ) to [ out=90, in=180 ]
        node [ node style multi ] { $\times$ }
        ( 6.5 - 0.1 , -\i * \width + 6.5 );
      }
    
      \draw[ arrow style ar ] ( 5 * \width + 0.5, 0.5 + 0.1 ) to [ out=90, in=180 ]
        node [ node style multi ] { $\times$ }
        ( 6.5 - 0.1 , -5 * \width + 6.5 );

      \node [ fill, white, inner sep=1 ] at ( 5., 6 + 0.5 ) {
        \begin{tabular}{|c|}
          \hline
          $f_{i,j}^0$ \\ 
          \hline
          $f_{i,j}^1$ \\ 
          \hline
          $f_{i,j}^2$ \\
          \hline
          $f_{i,j}^3$ \\
          \hline
        \end{tabular}
      };
      \node [ inner sep=0 ] (fvalbox) at ( 5., 6 + 0.5 ) {
        \begin{tabular}{|c|}
          \hline
          $f_{i,j}^0$ \\ 
          \hline
          $f_{i,j}^1$ \\ 
          \hline
          $f_{i,j}^2$ \\
          \hline
          $f_{i,j}^3$ \\
          \hline
        \end{tabular}
      };
      \draw [ thin ] (fvalbox.north east) -- ( 6.5, 7 ) ;
      \draw [ thin ] (fvalbox.south east) -- ( 6.5, 6 ) ;

      \node [ fill, white, inner sep=1 ] at ( 0.5, 1 + 0.5 ) {
        \begin{tabular}{|c|c|c|c|}
          \hline
          $\alpha_{i, j}^0$ & $\alpha_{i, j}^1$ & $\alpha_{i, j}^2$ & $\alpha_{i, j}^3$ \\
          \hline
        \end{tabular}
      };
      \node [ inner sep=0 ] (coefbox) at ( 0.5, 1 + 0.5 ) {
        \begin{tabular}{|c|c|c|c|}
          \hline
          $\alpha_{i, j}^0$ & $\alpha_{i, j}^1$ & $\alpha_{i, j}^2$ & $\alpha_{i, j}^3$ \\
          \hline
        \end{tabular}
      };
      \draw [ thin ] (coefbox.south west) -- ( 0, 0.5 ) ;
      \draw [ thin ] (coefbox.south east) -- ( 1, 0.5 ) ;

      \node [ inner sep=0 ] (fvalline) at ( 6.5 + 2.5, 7 + 1.25 ) {
        \begin{tabular}{|c|c|c|c|c|}
          \hline
          $f_{i,j}^0$ & $f_{i,j+1}^0$ & $f_{i,j+2}^0$ & $\ldots$ & $f_{i,j+\Ntheta-1}^0$ \\
          \hline
        \end{tabular}
      };

      \draw [ thin ] (fvalline.south west) -- ( 6.5, 7 ) ;
      \draw [ thin ] (fvalline.south east) -- ( 6.5 + 3.5, 7 ) ;

      \foreach \i in {0,1,...,2} {
        \emptyhcell{ 6.5 + \i * 0.5 }{ 0 }{ 0.5 }{ 0.5 }
      }
      \dothcell{ 6.5 + 3 * 0.5 }{ 0 }{ 3 * 0.5 }{ 0.5 }
      \emptyhcell{ 6.5 + 6 * 0.5 }{ 0 }{ 0.5 }{ 0.5 }

      \draw (           6.5 * \width, - 0.1 ) -- (           6.5 * \width, - 0.5 );
      \draw ( 7 * 0.5 + 6.5 * \width, - 0.1 ) -- ( 7 * 0.5 + 6.5 * \width, - 0.5 );
      \draw[ arrow style cote ] ( 6.5 * \width, - 0.25 ) to 
      node [ below, sloped, scale=0.9 ] {$\Ntheta$}
      ( 7 * 0.5 + 6.5 * \width, - 0.25 );

      \path (fvalline.north west) ++( 0, 0.8 ) node [ draw, scale = 1.2 ] {
      $ R_{i}(M_{final}) = R_{i}(M_{coef}) \times M_{fval} $
      };

    \end{tikzpicture}

    \caption{
      Illustration of a matrix-vector product between the weights
      (\texttt{contribution\_vector($i$)}\%val) and the
      value-derivatives of the input plane
      (\texttt{rhs\_product($i$)\%val}). One product computes
      $\Ntheta$ gyroaverage values.
    }
    \label{fig:sparse_matrix_vector_product}

  \end{center}
\end{figure}
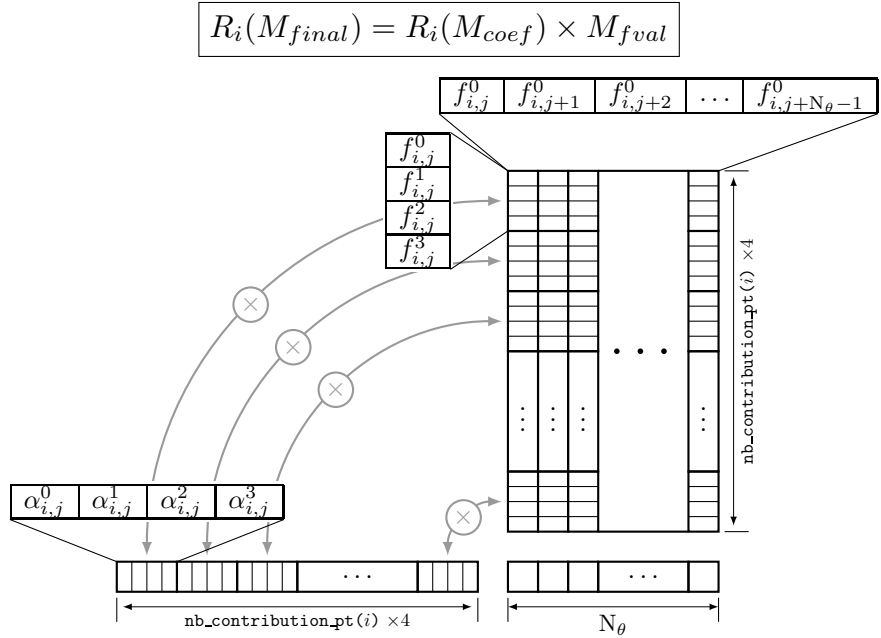

Algo.~\ref{algo:gyro_compute} which describes the main kernel of
gyroaverage can be decomposed into two parts: \emph{(i)} the computation
of the derivatives and \emph{(ii)} the computation of the product
between the weights and the \rhsval{} of the input plane. At
each call to \texttt{gyro\_compute}, the derivatives of the input
plane are computed and saved in a variable
denoted \texttt{rhs\_product}. This variable represents the matrix
$M_{fval}$. In this implementation, \texttt{rhs\_product} is
defined as follows:
\begin{listing}{1}
  type :: small_matrix
     real(8), dimension(:, :), pointer :: val
  end type small_matrix

  type(small_matrix), dimension(:), pointer :: rhs_product
\end{listing}

The variable \texttt{rhs\_product} is an array of length $(\Nr+1)$; it
contains elements of type \texttt{small\_matrix} which is a structure
containing simply a $2D$ pointer. The shape of a $2D$
array \texttt{rhs\_product(}$i$\texttt{)\%val} is \\
(\texttt{$4\ \times$ nb\_contribution\_pt($i$)},~$\Ntheta$). This
array is the matrix that contains the subset of values from matrix
$M_{fval}$ contributing to the gyroaverage of the
points ($i$,~$j\!=\!*$). Its first column
vector \\
\texttt{rhs\_product($i$)\%val(~:,$j=0$)}
contains the \rhsval{} retrieved from the first column vector
$\mathnormal{C_0}(M_{fval})$ which corresponds to the
position recorded in the
vector \texttt{contribution\_vector($i$)\%ind}. The other
columns \texttt{rhs\_product($i$)\%val(~:,$j\not=0$)}
are deduced thanks to a shift of
$j \times 4(\Nr+1)$ positions as one can see
in Fig.~\ref{fig:mfval_pattern}.

Fig.~\ref{fig:sparse_matrix_vector_product} illustrates the
matrix-vector product between the weights and the \rhsval{}. Here, the
\rhsval{} from a
vector \texttt{rhs\_product($i$)\%val(~:,$j$)}
are organized in such a way that their positions match with the correct weight
from the vector
\texttt{contribution\_vector($i$)\%val}. In this way,
the gyroaverage can be implemented as matrix-vector products which are
efficient operations on modern processors. Each matrix-vector
product computes the gyroaverage of $\Ntheta$ points. The gyroaverage
of the whole poloidal plane is achieved with $(\Nr+1)$ matrix-vector
products.

\subsubsection{\label{second_representation}Second representation of the matrix $M_{fval}$}

In this second approach, the variable \texttt{rhs\_product} is still
the representation of the matrix $M_{fval}$, but it is defined
as follows:
\begin{listing}{1}
  real(8), dimension(:), pointer :: rhs_product
\end{listing}

The \texttt{rhs\_product} variable is here a $1D$ array. It
corresponds exactly to the first column vector
$\mathnormal{C_0}(M_{fval})$. As said in
Section~\ref{pattern_matrix_fval}, this vector contains all the
\rhsval{} deduced from the input plane, so its length is
$4 (\Nr+1) \Ntheta$. One can see it as the linear flat representation of
the input plane.

\tikzset{ 
    arrow style ar/.style={ <->, >=latex, color=gray!80, thick },
    arrow style cote/.style={ <->, >=latex, very thin },
    node style text/.style={ below, sloped, scale=0.7 },
    node style multi/.style={ draw, circle, fill=white, midway, inner sep=1 },
}

\begin{figure}[t]
  \begin{center}
    \begin{tikzpicture}[ scale=0.8 ]
      \def\width{ 1 }

      \foreach \i in {0,1,...,2} {
        \hcell{ 0 + \i * \width }{ 0 }{ \width }{ 0.5 }
      }
      \dothcell{ 0 + 3 * \width }{ 0 }{ 2 * \width }{ 0.5 }
      \hcell{ 0 + 5 * \width }{ 0 }{ \width }{ 0.5 }

      \foreach \i in {0,1,...,2} {
        \vcell{ 0 + 6.5 * \width }{ 1 + \i * \width }{ 0.5 }{ \width }
      }
      \dotvcell{ 0 + 6.5 * \width }{ 1 + 3 * \width }{ 0.5 }{ 2 * \width }
      \vcell{ 0 + 6.5 * \width }{ 1 + 5 * \width }{ 0.5 }{ \width }

      \draw ( 0, 0 - 0.1 ) -- ( 0, -0.5 );
      \draw ( 6, 0 - 0.1 ) -- ( 6, -0.5 );
      \draw[ arrow style cote ] ( 0 + 0.1, -0.25 ) to 
      node [ node style text ] {\texttt{nb\_contribution\_pt($i$)} $\times$ 4}
      ( 6 - 0.1, -0.25 );

      \draw ( 7 + 0.1, 1 ) -- ( 7 + 0.5, 1 );
      \draw ( 7 + 0.1, 7 ) -- ( 7 + 0.5, 7 );
      \draw[ arrow style cote ] ( 7 + 0.25, 1 + 0.1 ) to 
      node [ node style text ] {$(\Nr+1) \times \Ntheta \times$ 4}
      ( 7 + 0.25, 7 - 0.1 );

      \draw[ arrow style ar ] ( 0.5, 0.5 + 0.1 ) to [ out=90, in=180 ] 
      node [ node style multi ] (prdt1) { $\times$ }
      ( 6.5 - 0.1 , 6.5 );
      \draw[ arrow style ar ] ( 0.5 + \width, 0.5 + 0.1 ) to [ out=90, in=180 ] 
      node [ node style multi ] (prdt2) { $\times$ }
      ( 6.5 - 0.1 , 5.5 );
      \draw[ arrow style ar ] ( 0.5 + 2*\width, 0.5 + 0.1 ) to [ out=90, in=180 ] 
      node [ node style multi ] (prdt3) { $\times$ }
      ( 6.5 - 0.1 , 2.5 );
      \draw[ arrow style ar ] ( 0.5 + 5*\width, 0.5 + 0.1 ) to [ out=90, in=200 ] 
      node [ node style multi ] (prdt4) { $\times$ }
      ( 6.5 - 0.1 , 4.5 );

      \node [ fill, white, inner sep=1 ] at ( 5., 6 + 0.5 ) {
        \begin{tabular}{|c|}
          \hline
          $f_{i,j}^0$ \\ 
          \hline
          $f_{i,j}^1$ \\ 
          \hline
          $f_{i,j}^2$ \\
          \hline
          $f_{i,j}^3$ \\
          \hline
        \end{tabular}
      };
      \node [ inner sep=0 ] (fvalbox) at ( 5., 6 + 0.5 ) {
        \begin{tabular}{|c|}
          \hline
          $f_{i,j}^0$ \\ 
          \hline
          $f_{i,j}^1$ \\ 
          \hline
          $f_{i,j}^2$ \\
          \hline
          $f_{i,j}^3$ \\
          \hline
        \end{tabular}
      };
      \draw [ thin ] (fvalbox.north east) -- ( 6.5, 7 ) ;
      \draw [ thin ] (fvalbox.south east) -- ( 6.5, 6 ) ;

      \node [ fill, white, inner sep=1 ] at ( 0.5, 1 + 0.5 ) {
        \begin{tabular}{|c|c|c|c|}
          \hline
          $\alpha_{i, j}^0$ & $\alpha_{i, j}^1$ & $\alpha_{i, j}^2$ & $\alpha_{i, j}^3$ \\
          \hline
        \end{tabular}
      };
      \node [ inner sep=0 ] (coefbox) at ( 0.5, 1 + 0.5 ) {
        \begin{tabular}{|c|c|c|c|}
          \hline
          $\alpha_{i, j}^0$ & $\alpha_{i, j}^1$ & $\alpha_{i, j}^2$ & $\alpha_{i, j}^3$ \\
          \hline
        \end{tabular}
      };
      \draw [ thin ] (coefbox.south west) -- ( 0, 0.5 ) ;
      \draw [ thin ] (coefbox.south east) -- ( 1, 0.5 ) ;

      \emptyhcell{ 6.5 + 0 * 0.5 }{ 0 }{ 0.5 }{ 0.5 }

      \path ( -1, 5.5 ) node [ rotate = 10, scale = 0.8 ] (matchlegend) {
      \texttt{contribution\_vector($i$)\%ind}
      };

      \draw [ dashed ] (matchlegend.south) edge[bend right]  (prdt1);
      \draw [ dashed ] (matchlegend.south) edge[bend right]  (prdt2);
      \draw [ dashed ] (matchlegend.south) edge[bend right]  (prdt3);
      \draw [ dashed ] (matchlegend.south) edge[bend right]  (prdt4);

      \path (fvalbox.north) ++( 0, 0.8 ) node [ draw, scale = 1.2 ] {
      $ M_{final}( i, j ) = R_{i}(M_{coef}) \times C_{j}(M_{fval}) $
      };

    \end{tikzpicture}

    \caption{
      Illustration of the sparse dot product between the weigths
      (\texttt{contribution\_vector($i$)\%val}) and the
      value-derivatives of the input plane
      (\texttt{rhs\_product}). The matching between these two arrays
      is done thanks to the position recorded
      in \texttt{contribution\_vector($i$)\%ind}.
    }
    \label{fig:sparse_dot_product}

  \end{center}
\end{figure}
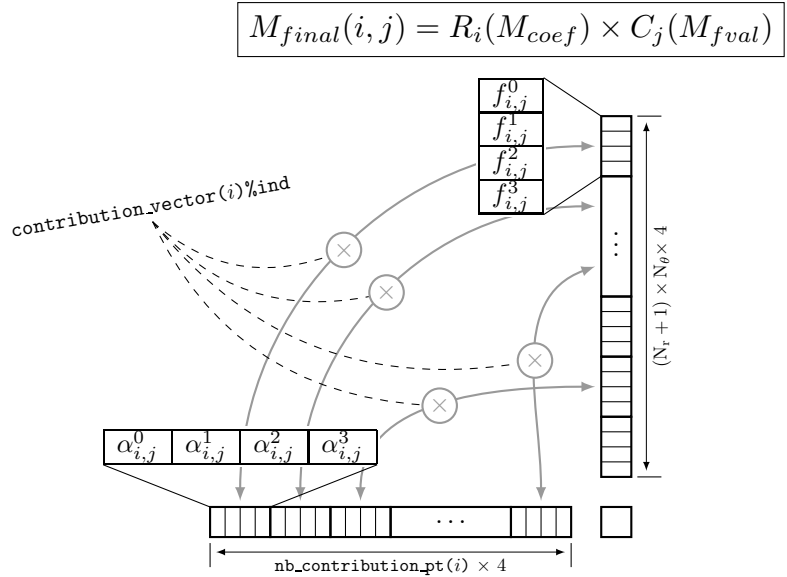

Fig.~\ref{fig:sparse_dot_product} shows the product between the
weights and the \rhsval{} accordingly to this representation.
The length of the vectors involved in this product does not match. The
association between weights and the \rhsval{} is done thanks
to the position recorded in the
vector \texttt{contribution\_vector($i$)\%ind}. This
association is graphically represented by the arrows. The gyroaverages
of points ($i$, $j \not= 0$) are computed thanks to a shift
of $4 j (\Nr+1)$ of the indexes
from \texttt{contribution\_vector($i$)\%ind}. This
operation will be called \emph{sparse dot product} in the following.

To achieve the gyroaverage of the whole poloidal plane, $(\Nr+1) \Ntheta$
sparse dot products are required.

\subsubsection{Benchmark to compare the data structures}

\begin{table}[t]
  \begin{center}
    \begin{tabular}{| c | c || c | c | c | c | c | c |}
      \hline

      \multicolumn{2}{| c ||}{$N_r = N_{\theta}$} & $32$ & $64$ & $128$ & $256$ & $512$ \\

      \hline
      \hline

      \multicolumn{1}{| c }{} & 
      RHS & 
      1,5 $\cdot 10^{ -3 }$ & 
      1,1 $\cdot 10^{ -2 }$ & 
      8,5 $\cdot 10^{ -2 }$ & 
      5,1 $\cdot 10^{ -1 }$ & 
      2,3 \\ 

      \cline{2-7}

      \multicolumn{1}{| c }{\textbf{Structure 1:}} & 
      Product &
      2,1 $\cdot 10^{ -4 }$ & 
      1,7 $\cdot 10^{ -3 }$ & 
      1,4 $\cdot 10^{ -2 }$ & 
      8,3 $\cdot 10^{ -2 }$ & 
      3,5 $\cdot 10^{ -1 }$ \\ 

      \cline{2-7}

      \multicolumn{1}{| c }{} & 
      Total & 
      1,7 $\cdot 10^{ -3 }$ & 
      1,3 $\cdot 10^{ -2 }$ & 
      9,9 $\cdot 10^{ -2 }$ & 
      5,9 $\cdot 10^{ -1 }$ & 
      2,7 \\ 

      \hline
      \hline

      \multicolumn{1}{| c }{} & 
      RHS & 
      1,7 $\cdot 10^{ -4 }$ & 
      6,4 $\cdot 10^{ -4 }$ & 
      2,5 $\cdot 10^{ -3 }$ & 
      1,0 $\cdot 10^{ -2 }$ & 
      4,2 $\cdot 10^{ -2 }$ \\ 

      \cline{2-7}

      \multicolumn{1}{| c }{\textbf{Structure 2:}} & 
      Product &
      7,9 $\cdot 10^{ -4 }$ & 
      5,7 $\cdot 10^{ -3 }$ & 
      4,6 $\cdot 10^{ -2 }$ & 
      2,7 $\cdot 10^{ -1 }$ & 
      1,2 \\ 

      \cline{2-7}

      \multicolumn{1}{| c }{} & 
      Total & 
      9,7 $\cdot 10^{ -4 }$ & 
      6,3 $\cdot 10^{ -3 }$ & 
      4,9 $\cdot 10^{ -2 }$ & 
      2,8 $\cdot 10^{ -1 }$ & 
      1,3 \\ 

      \hline

    \end{tabular}
  \end{center}
  \caption{
    Comparison of performance of the $2$ data structures on the
    analytical test case. Structure 1 corresponds to the first
    representation of $M_{fval}$ (described in
    section~\ref{first_representation}) and structure 2 to the second
    one (described in section~\ref{second_representation}).  The
    execution times are given in seconds. They result from the average
    over $100$ runs. ($\rho = 0.05$, $r_{min} = 0.1$,
    $r_{max} = 0.9$, $N = 32$).
  }
  \label{tab:benckmark_data_structure}
  \vspace*{-2em}
\end{table}

To compare the $2$ data structures introduced above, we used the
analytical test case as a benchmark. An execution of this program
consists in computing the gyroaverage of a plane initialized by a
chosen function and to compare the result with the associated
theoretical value, as it is described in
section~\ref{sec:analytical_test_case}. Let consider the
complexity \texttt{comp($\Nr$, $\Ntheta$)} representing the product between the
weights and the \rhsval{} as the number of multiplications
involved. In our case, it reads as follows:
\begin{equation*}
\texttt{comp($\Nr$, $\Ntheta$)} = \sum_{i = 0}^{\Nr} 4
                                  \times \texttt{nb\_contribution\_pt($i$)} \times \Ntheta .
\end{equation*}
This relation highlights the strong correlation between the number of
grid points and the time complexity of the gyroaverage.

To identify and compare the behavior of $2$ implementations of the
gyroaverage, a scan on the size of the poloidal plane has been done.
Tab.~\ref{tab:benckmark_data_structure} shows the execution time of
the gyroaverage for the different sizes of the input plane. The given
measurements have been obtained as the average over $100$ runs of the 
case presented in Section \ref{sec:analytical_test_case}. The time corresponding to the building of the
variable \texttt{rhs\_product} is given in lines "RHS" and the time
for the product between the weights and the \rhsval{} in lines
"Product".

The second data structure implementation is roughly twice faster than
the first one. If you pay attention to the details, the first data
structure is competitive during the "Product" step, but the time
needed to build the \texttt{rhs\_product} represents a large overhead.
As the second data structure gives shorter execution times, this is
the chosen implementation.

With this data structure, the accesses to the \rhsval{} are
not contiguous during the sparse dot product (see
Fig.~\ref{fig:sparse_dot_product}). This is due to the irregular
stencil required by the gyroaverage (see
Fig.~\ref{fig:gyroaverage_example}) and to the layout of the
variable \texttt{rhs\_product}. This constitutes a limit for this
approach from the performance point of view.

The second data structure is trickier to manipulate for the sparse dot
product as it is shown in Section~\ref{second_representation}, but it
allows us to have the control on the internal data distribution which is
the key point for the "layout optimization" (details in
Section~\ref{sec:layout}).

To achieve the gyroaverage of the whole plane, at each grid point, a
sparse dot product has to be done. In the implementation on the
current "compact vector optimization", the poloidal plan is covered
thanks to $2$ nested for-loop where $j$ is the index of the
internal loop. Geometrically, the plan is explored circle after
circle, from the smaller to the larger ones. This access pattern has
the property to extensively reuse the weights
(\texttt{contribution\_vector}). Although this property, it
constitutes a limit for performance and represents the key point of
the next optimization. We will refer to this access pattern to cover
the plane as the \emph{mesh\_path} in the next sections.

\subsection{\label{sec:blocking}Blocking optimization}

As said previously, to achieve the gyroaverage of the whole poloidal
plane, a sparse dot product must be done on every grid point. The
"blocking optimization" extends the "compact vector optimization"
implementation. This optimization is focused on the improvement of the
access pattern \emph{mesh\_path} to the input plan inspired by the
well-known blocking/tiling technique \cite{Wolf:1991:DLO:113446.113449}.

To improve the performance, the idea here is to follow a path over the
grid points \emph{mesh\_path} which reduces the frequencies of cache
misses during the computation. For any mesh point ($i$, $j$),
the computation of the gyroaverage required data
from \texttt{contribution\_vector($i$)}
and \texttt{rhs\_product}. As said in~\ref{sec:compact_vector}, doing
the gyroaverage circle after circle ensures a good reuse of the
weights \texttt{contribution\_vector(}$i$\texttt{)}, but this is not
the case for \texttt{rhs\_product}. The access to its data is
generally not contiguous and differs for $2$ distinct mesh
points. However, as the data are retrieved by cache line, the
cache \textit{hit rate} depends of
the \emph{mesh\_path}\footnote{http://gameprogrammingpatterns.com/data-locality.html}. The
goal achieved in this optimization is to increase the reuse of the
data available in cache along the \emph{mesh\_path}.

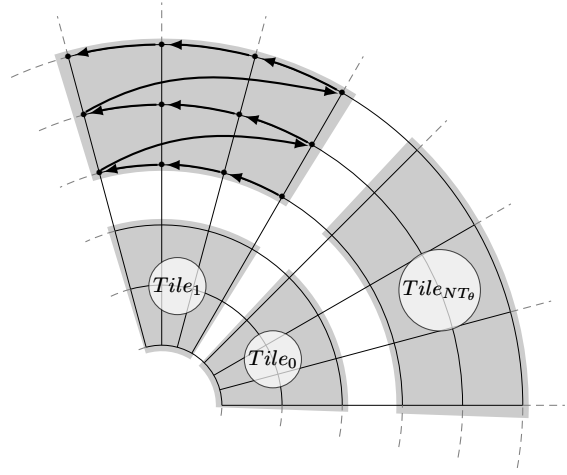
\begin{figure}[t]
  \begin{center}
    \begin{tikzpicture}[ scale=0.8 ]

      \begin{scope}[
          rotate = +22.5,
        ]
        \fill [ color=gray!40 ] ( -22.5 -2 : 3 +0.1 )
        arc (  -22.5 -2 : +22.5 +2 : 3 +0.1 )
        --  ( +22.5 +2 : 1 -0.1 )
        arc ( +22.5 +2 : -22.5 -2 : 1 -0.1 )
        -- cycle ;
      \end{scope}

      \begin{scope}[
          rotate = +22.5 + 60,
        ]
        \fill [ color=gray!40 ] ( -22.5 -2 : 3 +0.1 )
        arc (  -22.5 -2 : +22.5 +2 : 3 +0.1 )
        --  ( +22.5 +2 : 1 -0.1 )
        arc ( +22.5 +2 : -22.5 -2 : 1 -0.1 )
        -- cycle ;
      \end{scope}

      \begin{scope}[
          rotate = +22.5,
        ]
        \fill [ color=gray!40 ] ( -22.5 -2 : 6 +0.1 )
        arc (  -22.5 -2 : +22.5 +2 : 6 +0.1 )
        --  ( +22.5 +2 : 4 -0.1 )
        arc ( +22.5 +2 : -22.5 -2 : 4 -0.1 )
        -- cycle ;
      \end{scope}

      \begin{scope}[
          rotate = +22.5 + 60,
        ]
        \fill [ color=gray!40 ] ( -22.5 -2 : 6 +0.1 )
        arc (  -22.5 -2 : +22.5 +2 : 6 +0.1 )
        --  ( +22.5 +2 : 4 -0.1 )
        arc ( +22.5 +2 : -22.5 -2 : 4 -0.1 )
        -- cycle ;
      \end{scope}

      \foreach \i in {0, ..., 6 } {
        \draw [ very thin ] ( \i, 0 ) arc ( 0 : 105 : \i );
      }

      \foreach \i in {0, ..., 7 } {
        \draw [ very thin ] ( 0 + \i * 15 : 1 ) -- ( 0 + \i * 15 : 6 );
      }

      \foreach \i in {1, ..., 6 } {
        \draw [ densely dashed, thin, color = gray!100 ] 
        ( -10 : \i ) arc ( -10 : 0 : \i );
        \draw [ densely dashed, thin, color = gray!100 ] 
        ( 105 : \i ) arc ( 105 : 105 + 10 : \i );
      }

      \foreach \i in { 0, ..., 7 } {
        \draw [ densely dashed, thin, color = gray!100 ] 
        ( 0 + \i * 15 : 6 ) -- ( 0 + \i * 15 : 6.7 );
      }

      \foreach \i in { 4, ..., 6 } {
        \foreach \j in { 4, ..., 6 } {
          \draw [ ->, >=latex, thick ] ( 0 + \i*15 +1 : \j )
          arc ( 0 + \i*15 +1 : 15 + \i*15 -1 : \j );
        }
      }
      \draw [ ->, >=latex, thick ] ( 0 + 7*15 -0.5: 4 +0.04) to [out = 30,
        in = 170 ] ( 0 + 4*15 +0.5: 5 -0.04);
      \draw [ ->, >=latex, thick ] ( 0 + 7*15 -0.5: 5 +0.04) to [out = 30,
        in = 170 ] ( 0 + 4*15 +0.5: 6 -0.04);

      \foreach \i in { 4, ..., 7 } {
        \foreach \j in { 4, ..., 6 } {
          \node [ scale = 0.55 ] at ( 0 + \i * 15 : \j ) { $\bullet$ } ;
        }
      }

      \node [ draw, circle, fill=white, opacity=0.7, 
        inner sep=1pt, scale = 0.8 ] at ( 45 / 2 : 2 ) { $Tile_0$ } ;
      \node [ scale = 0.8 ] at ( 45 / 2 : 2 ) { $Tile_0$ } ;

      \node [ draw, circle, fill=white, opacity=0.7, 
        inner sep=1pt, scale = 0.8 ] at ( 82.5 : 2 ) { $Tile_1$ } ;
      \node [ scale = 0.8 ] at ( 82.5 : 2 ) { $Tile_1$ } ;

      \node [ draw, circle, fill=white, opacity=0.7, 
        inner sep=1pt, scale = 0.8 ] at ( 45 / 2 : 5 ) { $Tile_{NT_{\theta}}$ } ;
      \node [ scale = 0.8 ] at ( 45 / 2 : 5 ) { $Tile_{NT_{\theta}}$ } ;

    \end{tikzpicture}

    \caption{
      Sample of the tiling technique over a part of the poloidal
      mesh. The tiles which partition the mesh are traveled in
      ascending order. The nested arrows inside a tile show the
      underlying order.
    }
    \label{tiling_path}

  \end{center}
\end{figure}

Instead of covering the poloidal plan circle after circle, the plan is
covered using small $2D$ tiles in this version. These tiles are of
size $BLOCK_r \times BLOCK_\theta$ and are numbered following the
poloidal direction. Fig.~\ref{tiling_path} illustrates the tiling
decomposition of a part of the poloidal mesh. On this sample, $BLOCK_r
= 3$ and $BLOCK_\theta = 4$. In practice, the setting of $BLOCK_r$ and
$BLOCK_\theta$ are tuned depending on benchmarks performed on each production
machine.

The poloidal plane is partitioned into $NT_r \times NT_\theta$ tiles.
The tiles are numbered along the poloidal direction, from the center
to the outside of the plane. This \emph{mesh\_path} cut in tiles
improves the temporal locality of the data from the
array \texttt{rhs\_product}. It increases the cache \textit{hit rate}
and so decreases the execution time.

\subsection{\label{sec:layout}Layout optimization}

As for the "blocking optimization", the "compact vector optimization"
is the starting point of this implementation. Especially,
the \emph{mesh\_path} remains the same in the present
implementation. Here, this optimization focuses on the order of the
data in the array \texttt{rhs\_product}. For a
given \emph{mesh\_path}, the arrangement of these values has a
strong impact on the execution time. We will refer to this
arrangement of data as the \emph{layout} of \texttt{rhs\_product}.

The optimization of the layout is done for a given \emph{mesh\_path}
over the input plane. The access history to the data
of \texttt{rhs\_product} depends on the \emph{mesh\_path}. The idea
of the optimization presented here is to permute the elements
of \texttt{rhs\_product}, so change its layout, in order to increase
the cache \textit{hit ratio} along the \emph{mesh\_path}.

The \emph{mesh\_path} must be seen as an input parameter of the
problem. Depending on the gyroradius $\rho$, the number of cells
$\Nr \times \Ntheta$, the number of points on the Larmor circle $N$
and the \emph{mesh\_path}, the access history to the
array \texttt{rhs\_product} is known. To determine how to change the
layout, we need to introduce a metric of temporal distance of access
of its data.

\begin{table}[htb]
  \centering{
    \begin{tabular}{| c | l | c |}
      \hline
      contribution point & list of temporal indexes             & temporal\_average \\
      \hline
      \hline
                  (0, 0) & [0, 1, 8, 9, 10, 17]                 &  7.0     \\
                  (0, 1) & [0, 1, 8, 9, 10, 17, 18, 19, 26]     & 12.0     \\
                  (0, 2) & [9, 10, 17, 18, 19, 26, 27, 28, 35]  & 21.0     \\
                  (0, 3) & [18, 19, 26, 27, 28, 35, 36, 37, 44] & 30.0     \\
                $\vdots$ & $\vdots$                             & $\vdots$ \\
                  (1, 0) & [0, 1, 2, 9, 10, 11]                 &  5.0     \\
                $\vdots$ & $\vdots$                             & $\vdots$ \\
      \hline
    \end{tabular}
 
    \caption{
    For each mesh point, list of the indexes corresponding to their use
    along the \emph{mesh\_path} and average of this
    temporal indexes.
    }
 
    \label{tab:temporal_index}
  }
\end{table}

To compute the gyroaverage at a point ( $i$, $j$ ), the
\rhsval{} values of its neighborhood are required. The points of this
neighborhood are the contribution points to the gyroaverage at the
point ( $i$, $j$ ). To achieve the gyroaverage of a whole poloidal
plan, we compute the gyroaverage of each mesh point one after an other.
Let the \emph{temporal index} be the
index of a grid point along the \emph{mesh\_path}. For each point of
the plane, we have listed the temporal indexes corresponding to the
gyroaverages in which it is involved. 

Tab.~\ref{tab:temporal_index}
gives a sample of the association between the mesh points and their
temporal indexes. For instance, thanks to this table, one knows when
the point \texttt{(0,0)} is used along the \emph{mesh\_path} during
the sparse dot products: 0, 1, 8, 9, 10 and 17. The second information
given by this table is the average of the temporal indexes for each
contribution point, which is denoted \emph{temporal\_average}. For example,
the \emph{temporal\_average} of the point \texttt{(0,0)} is computed 
as follows:
$$
\text{temporal\_average}( \mathtt{(0, 0)} ) = \frac{0 + 1 + 8 + 9 + 10 + 17}{6} = 7.
$$
Let \texttt{pt1} a mesh point. If its \emph{temporal\_average} is rather low, this means \texttt{pt1} contributes rather in the beginning of the \emph{mesh\_path}. On the opposite, if the \emph{temporal\_average} of a mesh point \texttt{pt2} is rather high compared to the other temporal averages, this means \texttt{pt2} contributes rather in the end of the \emph{mesh\_path}.

The heuristic implemented in this optimization is to sort the elements
of \texttt{rhs\_product} accordingly to their \emph{temporal\_average}. 
For instance, from the information available on
Tab.~\ref{tab:temporal_index} the resulting partial data layout
of \texttt{rhs\_product} is: [ (1,~0), (0,~0), (0,~1), (0,~2), (0,~3),
$\ldots$ ]. The main idea is to keep accessed data between $2$
successive sparse dot products as near as possible in order to
minimize in average the number of cache misses.

In practice, the computation of the new layout of \texttt{rhs\_product}
is done during the initialization step. This implies that the
\emph{mesh\_path} over the plane must be known at this step.
The matching between the weights
(\texttt{contribution\_vector()\%val}) and the \rhsval{}
(\texttt{rhs\_product}) during a sparse dot product is done thanks to
the indexes retrieved from the
array \texttt{contribution\_vector()\%ind}. To ensure the validity
of the computation, the indexes recorded in
array \texttt{contribution\_vector()\%ind} must be updated
accordingly to this new layout.

\section{\label{perf_results}Performance results in \gy}

\subsection{Benchmarking small cases}

After integration of the different versions of the gyroaverage
operator in \gy code~\cite{grandgirard:cea-01153011}, different runs have been done to compare their
effectiveness. To keep a reasonably short execution time, the
simulations are composed of only four time steps. These tests were
done on the \hel\footnote{IFERC-CSC HELIOS super computer in
Rokkasho-Japan, http://www.top500.org/system/177449}
machine. Computation nodes used are equipped with two Intel Xeon
E5-2450 2.10GHz processors, so 16 cores a node. On
Tab.~\ref{tab:perf_res16}, the different test cases have been
performed over the two following meshes~\eqref{eq:mesh2} and
\eqref{eq:mesh3}:\\[-.2cm]
\begin{equation}
N_r\!=\!256,\,N_\theta\!=\!256,\,N_\varphi\!=\!32,\,N_{v_\parallel}\!=\!16,\,N_{\mu}\!=\!4\,
\label{eq:mesh2}
\end{equation}
\begin{equation}
N_r\!=\!512,\,N_\theta\!=\!512,\,N_\varphi\!=\!32,\,N_{v_\parallel}\!=\!16,\,N_{\mu}\!=\!4 .
\label{eq:mesh3}
\end{equation}

To highlight the difference of performance between the different
gyroaverage implementations, the size of the poloidal plane is
multiplied by 4 between the two meshes. For these runs, every
gyroaverage based of Hermite interpolation has $N\!=\!16$ quadrature
points on the integration circle. This number of integration points
ensures a good presicion of the results as you can see on
Tab.~\ref{tab:growth_rates}. The fluctuation of execution times
(several percents typically, but rare events can lead to 10\% or 20\%)
due to shared resources such as network and parallel file system is
avoided using a specific configuration. Only one computation node of
\hel is used for each run in order not to share the network. Also, the
writing and reading on the parallel file system is reduced to the
minimum. The benchmark configuration of one run is the following: $16$
\mpi processes composed of only $1$ \omp thread. This benchmark 
corresponds to an execution of only four time steps of \gy{} of a non physical 
case. This quite artificial setting will be revised in the next 
Subsection dedicated to production runs.

\begin{table*}[!h]
  \begin{center}
\scalebox{0.9}{
    \begin{tabular}{|l|c|c|*{2}{c|}*{2}{c|}c|}
      \cline{3-8}

      \multicolumn{2}{c|}{}                   &
      \multicolumn{6}{c|}{Gyroaverage method} \\

      \cline{3-8}

      \multicolumn{2}{c|}{}        & 
      Pad\'e                       & 
      \multicolumn{4}{c|}{Hermite} & 
      \texttt{Disable}             \\

      \cline{4-7}

      \multicolumn{2}{c|}{}       & 
      \multicolumn{1}{c|}{}       & 
      Initial                      &
      Compact vector               &
      Layout                       &
      Blocking                     &
                                  \\
      \hline
      \hline

%
%
%
      \multicolumn{1}{|c}{\textbf{Mesh~\eqref{eq:mesh2}}:} &
      Total execution time (sec.)  &
      110,20 & 
      150,60 & 
      146,35 & 
      132,91 & 
      131,39 & 
      \texttt{106,18} \\ 

      \cline{2-8}

      \multicolumn{1}{|c}{} &
      Percentage of gyroavg. execution &
      3.8 \% &
      41.8 \% &
      37.8 \% &
      25.2 \% &
      23.7 \% &
      -- \\
      \hline
      \hline

      \multicolumn{1}{|c}{\textbf{Mesh~\eqref{eq:mesh3}}:} &
      Total execution time (sec.)  &
      464,62 & 
      629,46 & 
      627,80 & 
      557,49 & 
      550,38 & 
      \texttt{433,35} \\ 

      \cline{2-8}

      \multicolumn{1}{|c}{} &
      Percentage of gyroavg. execution &
      7.2 \% &
      45.3 \% &
      44.9 \% &
      28.6 \% &
      27.0 \% &
      -- \\
      \hline
      \hline
    \end{tabular}
}
  \end{center}
  \caption{ Execution time and percentage over the total time of the
    gyroaverage operator for the different versions (for two different
    mesh sizes of poloidal plane with $N=16$ interpolation points).  }
  \label{tab:perf_res16}
\end{table*}

Tab.~\ref{tab:perf_res16} shows some results concerning the Pad\'e
approximation compared to $4$ versions of the Hermite implementation:
original implementation; compact vector (described in
Section~\ref{sec:compact_vector}), blocking
(Section~\ref{sec:blocking}) and layout (Section~\ref{sec:layout})
optimizations.  Also, the \gy code can be executed without any
gyroaverage at all (solution named \texttt{Disable} gives invalid
results of course but consitutes a reference execution time). This last option allows us to determine
accurately the fraction of the execution time corresponding to the
gyroaverage operator over the total execution time for the other
cases.

As expected, the computation time of the gyroaverage operator grows
along with the size of the poloidal mesh. The \textit{compact vector
  optimization} which were an intermediate step during the
optimization process is already a little bit faster than the original
implementation. However, the percentage of total execution time
dedicated to gyroaverage operator is about $6$ to $10$ times higher
than for the Pad\'e approximation. The \textit{layout optimization}
offers a great speed up compared to this previous step. On the biggest
mesh, it reduces the execution by several tens of percents compared to
the \textit{compact vector optimization}. This demonstrates the big
impact of the data layout in memory.

The best result is obtained with the \textit{blocking optimization}
which reduces the accumulated gyroaverage cost by 40\% compared to the
\textit{initial} version. Nevertheless, it is still approximately 4
times more costly than the Pad\'e approximation on the largest
mesh. This is partly due to the number $N$ of points used on the
integration circle.

\begin{table*}[!h]
  \begin{center}
\scalebox{0.9}{
    \begin{tabular}{|l|c|c|*{2}{c|}*{2}{c|}c|}
      \cline{3-8}

      \multicolumn{2}{c|}{}                   &
      \multicolumn{6}{c|}{Gyroaverage method} \\

      \cline{3-8}

      \multicolumn{2}{c|}{}        & 
      Pad\'e                       & 
      \multicolumn{4}{c|}{Hermite} & 
      \texttt{Disable}             \\

      \cline{4-7}

      \multicolumn{2}{c|}{}       & 
      \multicolumn{1}{c|}{}       & 
      Initial                      &
      Compact vector               &
      Layout                       &
      Blocking                     &
                                  \\
      \hline
      \hline
%
%

      \multicolumn{1}{|c}{\textbf{Mesh~\eqref{eq:mesh2}}:} &
      Total execution time (sec.)  &
      110,00 & 
      131,42 & 
      129,11 & 
      123,12 & 
      120,47 & 
      \texttt{105,60} \\ 

      \cline{2-8}

      \multicolumn{1}{|c}{} &
      Percentage of gyroavg. execution &
      4.2 \% &
      24.4 \% &
      22.3 \% &
      16.6 \% &
      14.1 \% &
      -- \\
      \hline
      \hline

      \multicolumn{1}{|c}{\textbf{Mesh~\eqref{eq:mesh3}}:} &
      Total execution time (sec.)  &
      464,20 & 
      544,07 & 
      538,38 & 
      525,21 & 
      510,44 & 
      \texttt{433,49} \\ 

      \cline{2-8}

      \multicolumn{1}{|c}{} &
      Percentage of gyroavg. execution &
      7.1 \% &
      25.5 \% &
      24.2 \% &
      21.2 \% &
      17.8 \% &
      -- \\
      \hline
      \hline
    \end{tabular}
}
  \end{center}
  \caption{
Execution time and percentage over the total time of the
    gyroaverage operator for the different versions (for two different
    mesh sizes of poloidal plane with $N=8$ interpolation points).
  }
  \label{tab:perf_res8}
\end{table*}

On Tab.~\ref{tab:perf_res8}, the same simulations have been run with
less quadrature points, so $N\!=\!8$.  As expected, reducing $N$
induces smaller computation costs and decreases the execution
times. With twice less quadrature points, the gyroaverage based on
Hermite (\textit{blocking}) reduces its execution time by 40\%. Also
the gap between the \textit{blocking optimization} and the Pad\'e
approximation is reduced. However the drawback of taking less
interpolation points leads to a subtle degradation of the quality of the
gyroaverage operator. The trade-off between precision and computation
will be discussed in the Subsection below.

\subsection{Benchmarking a real-life case }
\newcommand{\krhoi}{k_{\perp}{\rho}_i}

Let us consider a regular domain size and a set of commonly used input
parameters to analyze the performance of the new gyroaverage operator
on a real-life case. The execution times will be a little bit tainted
by network and parallel file system usage, because they are shared by
several users.  We have launched the simulations several times to
ensure the execution times we get are reproducible.  We focus on a
Cyclone DIII-D base case, with a most unstable mode at
$(m,n)=(14,-10)$ and the following parameters: $\mu_{min}=0.143$,
$\mu_{max}=7.$, radial size $a=100$ (described in \cite{Steiner2014},
p. 10). The domain size is $N_r=256$, $N_\theta=256$, $N_\varphi=64$,
$N_{v_{\parallel}}=48$, $N_{\mu}=8$. This kind of typical benchmark
had already been performed several years ago to validate the \gy
code. In order to check the good behavior of the simulation, a
classical verification procedure is to focus on the linear phase and
to extract the growth rate of the most unstable mode.

In the literature (see for example \cite{broemstrup}, p.  39-40), some
theoretical and applied works have pointed out that taking $N\!=\!8$
points to discretize the gyroaverage operator is often sufficient.  It
has also been shown that this choice is already better than Pad\'{e}
to approximate the Bessel function ${J_0}(\krhoi)$ on the interval
$\krhoi\in[0,5]$.  Working with $N\!=\!16$ points to evaluate the
gyroaverage extends the capability of the Hermite interpolation which
is now able to accurately approximate the Bessel function on a larger
interval $\krhoi\in[0,10]$. Fig.~\ref{fig:J0} also corroborates this
fact by an illustration on the specific function $f_{8,8}$.

\begin{table*}[!h]
  \begin{center}
\scalebox{0.9}{
\def\arraystretch{1.3}%
    \begin{tabular}{|c || c|c|c|c|c|c|c|}
      \hline
      Method & Pad\'{e} & \multicolumn{6}{c|}{Hermite} \\
      \cline{3-8}
      & & N=2 & N=3 & N=4 & N=6 & N=8 & N=16\\

      \hline

      Growth rate \hspace*{.1cm}$\gamma$ & .10436 & .12246 & .09420 & .09625 & .09643 & .09644 & .09644\\

      \hline
    \end{tabular}
}
  \end{center}

\vspace*{0.3cm}
  \caption{Linear growth rate for the different versions of the
    gyroaverage operator on a Cyclone test case (normalized growth rate 
    as Fig. 1 of~\cite{dimits}).}
  \label{tab:growth_rates}
\end{table*}

On Tab.~\ref{tab:growth_rates}, the linear growth rates (normalized
as in\,\cite{dimits}, Fig.~1) observed in the described use case are
presented. They characterize the behavior of all the main components
of the code during the linear phase. One can see that the values given
by Hermite method for $N\!=\!6$, $N\!=\!8$, $N\!=\!16$ are almost
equal, which is a good result that shows that simulations are
well converged. The asymptotic value is reached quickly as $N$ grows.
However, for $N\!=\!2$, $N\!=\!3$ and for the Pad\'{e} method,
the growth rate is not well recovered.  Practically, $N\!=\!8$ should
be taken for production runs and this is a good cost-quality balance. The
$N\!=\!16$ configuration will be useful also in specific cases where
great accuracy is wanted by physicists (whenever particles with
$\krhoi>5$ play a major role).


\begin{table*}[!h]
  \begin{center}
\scalebox{0.82}{
\def\arraystretch{1.3}%
    \begin{tabular}{|c || c|c|c|c|c|c|c|}
      \hline
      & & \multicolumn{6}{c|}{Hermite} \\
      \cline{3-8}
      Code part $\backslash$ Method & Pad\'{e} & \multicolumn{3}{c|}{N=8} & \multicolumn{3}{c|}{N=16}\\
      \cline{3-8}
      & & Initial & Layout & Blocking & Initial & Layout & Blocking \\
      \hline
      Field solver &  28s (0\%) &  32s (+14\%) &  31s (+10\%) &  31s (+11\%) &  35s (+27\%) &  32s (+13\%) &  32s (+16\%) \\
      \hline         
      Diagnostics  &  96s (0\%) & 123s (+29\%) & 108s (+12\%) & 110s (+15\%) & 147s (+54\%) & 114s (+19\%) & 120s (+26\%) \\
      \hline         
      Total        & 629s (0\%) & 662s (+5.3\%)& 659s (+4.8\%)& 651s (+3.6\%)& 689s (+9.6\%)& 670s (+6.6\%)& 666s (+5.8\%)\\
      \hline    
    \end{tabular}
}
  \end{center}

\vspace*{0.3cm}
  \caption{
    Cumulative execution time for some parts of the code that are
    impacted by gyroaverage computation costs. Timings are given for a
    run of 60 time steps using 512 cores. In parentheses, the
    percentage of extra time compared to Pad\'{e} reference time is
    given.
  }
  \label{tab:time_real_case}
\end{table*}

Several execution times are shown in Tab.~\ref{tab:time_real_case}
that figure out the behavior of the Cyclone case with different
versions of the gyroaverage operator. One can see that the
\textit{Diagnostics} and \textit{Field solver} parts which use the
gyroaverage operator are impacted by the chosen method. The
improved versions (Layout and Blocking) divides by a factor 2 the
overheads due to the Hermite interpolation method for the
\textit{Diagnostics}. On the overall total execution time, the best
methods add 4\% of extra time calculation for $N=8$ points and only 6\%
for $N=16$. This overhead is fully acceptable, as the accuracy of numerical
results are greatly improved. Furthermore, the new versions described
in this paper diminish the execution time of the gyroaverage by at
least 40\% compared to the initial method (Tab.~\ref{tab:perf_res8}).

\section{\label{conclusion}Conclusion}

To achieve the optimization of the gyroaverage operator, $4$ steps
have been achieved: \emph{(i)} derive numerical approximation of the
gyroaverage operator,
\emph{(ii)} describe formally the operator thanks to a matrix representation,
\emph{(iii)} explore and evaluate different optimization techniques, and
\emph{(iv)} integrate and validate these optimizations in the \gy code.
The current best implementation with Hermite interpolation is twice faster than the initial
one. This improvement, but also the enhanced accuracy of the
gyroaverage based on Hermite interpolation foster us to use the new
solution in production runs instead of the Pad\'e approximation.

The Pad\'{e} gyroaverage implementation required a whole poloidal cut
as input, due to a Fourier transform in $\theta$ direction and to a
finite difference discretization along $r$ direction. This involves sometimes
collective communications to redistribute the data before applying the
gyroaverage, because the main parallel domain decomposition is along
$r$, $\theta$ and $\mu$ directions in \gy. Historically, this was a
hard constraint because Pad\'{e} was the single solution available in
Gysela for gyroaveraging. The new gyroaverage based on interpolation
methods circumvents this issue. Parallelizing among
several \mpi processes will, from now on, be greatly eased.
Additionally this new way to compute the gyroaverage allows us to
consider easily its extension over general geometry meshes (not only 
polar planes).


\bibliographystyle{plain}
\bibliography{gyro_ref}

\begin{thebibliography}{10}

\bibitem{Bigot_ESAIM2013}
J.~Bigot, V.~Grandgirard, G.~Latu, Ch. Passeron, F.~Rozar, and O.~Thomine.
\newblock Scaling gysela code beyond 32k-cores on bluegene/q.
\newblock In {\em CEMRACS 2012}, volume~43 of {\em ESAIM: Proc.}, pages
  117--135, Luminy, France, 2013.

\bibitem{broemstrup}
I.~Broemstrup.
\newblock {\em Advanced Lagrangian Simulation Algorithms for Magnetized Plasma
  Turbulence}.
\newblock PhD thesis, Univ. of Maryland, 2008.

\bibitem{cms}
N.~Crouseilles, M.~Mehrenberger, and H.~Sellama.
\newblock Numerical solution of the gyroaverage operator for the finite
  gyroradius guiding-center model.
\newblock {\em Commun. Comput. Phys.}, 8, 2010.

\bibitem{dimits}
A.~M.~Dimits et~al.
\newblock Comparisons and physics basis of tokamak transport models and
  turbulence simulations.
\newblock {\em Physics of Plasmas}, 7(3):969--983, 2000.

\bibitem{grandgirard:cea-01153011}
V.~Grandgirard, J.~Abiteboul, J.~Bigot, T.~Cartier-Michaud, N.~Crouseilles, Ch.
  Erhlacher, D.~Esteve, G.~Dif-Pradalier, X.~Garbet, Ph. Ghendrih, G.~Latu,
  M.~Mehrenberger, C.~Norscini, Ch. Passeron, F.~Rozar, Y.~Sarazin,
  A.~Strugarek, E.~Sonnendr{\"u}cker, and D.~Zarzoso.
\newblock {A 5D gyrokinetic full-f global semi-lagrangian code for flux-driven
  ion turbulence simulations}.
\newblock working paper or preprint, July 2015.

\bibitem{Grandgirard_JOCP2006}
V.~Grandgirard, M.~Brunetti, P.~Bertrand, N.~Besse, X.~Garbet, Ph. Ghendrih,
  G.~Manfredi, Y.~Sarazin, O.~Sauter, E.~Sonnendr{\"u}cker, J.~Vaclavik, and
  L.~Villard.
\newblock A drift-kinetic semi-lagrangian 4d code for ion turbulence
  simulation.
\newblock {\em Journal of Computational Physics}, 217(2):395 -- 423, 2006.

\bibitem{Grandgirard_PPCF2007}
V.~Grandgirard, Y.~Sarazin, P.~Angelino, A.~Bottino, N.~Crouseilles, G.~Darmet,
  G.~Dif-Pradalier, X.~Garbet, Ph. Ghendrih, S.~Jolliet, G.~Latu,
  E.~Sonnendr{\"u}cker, and L.~Villard.
\newblock Global full-f gyrokinetic simulations of plasma turbulence.
\newblock {\em Plasma Physics and Controlled Fusion}, 49(12B):B173, 2007.

\bibitem{krehbessel}
M.~Kreh.
\newblock Bessel functions.
\newblock {\em Lecture Notes, Penn State-G{\"o}ttingen Summer School on Number
  Theory}, 2012.

\bibitem{Latu_PPAM2012}
G.~Latu, V.~Grandgirard, N.~Crouseilles, and G.~Dif-Pradalier.
\newblock Scalable quasineutral solver for gyrokinetic simulation.
\newblock Research Report RR-7611, INRIA, May 2012.

\bibitem{Li2005}
J.~Li.
\newblock General explicit difference formulas for numerical differentiation.
\newblock {\em Journal of Computational and Applied Mathematics},
  183(1):29--52, 2005.

\bibitem{Steiner2014}
C.~Steiner, M.~Mehrenberger, N.~Crouseilles, V.~Grandgirard, G.~Latu, and
  F.~Rozar.
\newblock Gyroaverage operator for a polar mesh.
\newblock {\em The European Physical Journal D}, 69(1), 2015.

\bibitem{Wolf:1991:DLO:113446.113449}
M.~E. Wolf and M.~S. Lam.
\newblock A data locality optimizing algorithm.
\newblock {\em SIGPLAN Not.}, 26(6):30--44, May 1991.

\end{thebibliography}

\end{document}